\documentclass[11pt]{article}
\usepackage{amsmath,amsthm,amssymb,cite,enumerate}
\usepackage[a4paper,left=0.5in,right=1in]{geometry}
\usepackage[usenames]{color}
\usepackage{graphicx}
\usepackage{epsfig}
\usepackage{dcolumn}% Align table columns on decimal point
\usepackage{bm}% bold math

\begin{document}

\def \d {{\rm d}}
\def \q {Q}
\def \etta {F} %\def \etta {\eta}
\def \t {{\Theta}}
\def \k {{\kappa}}
\def \l {{\lambda}}
\def \s {{\sigma}}

\def \tv {{\tilde v}}
\def \tz {{\tilde z}}

\def \P {{p_\lambda}}
\def \Q {{p_\nu}}
\def \eigen {{\mathcal N}}
\def \M {{\mathcal M}}
\def \bm #1 {\mbox{\boldmath{$m^{(#1)}$}}}

\def \bF {\mbox{\boldmath{$F$}}}
\def \bk {\mbox{\boldmath{$k$}}}
\def \bl {\mbox{\boldmath{$l$}}}
\def \bbm {\mbox{\boldmath{$m$}}}
\def \tbbm {\mbox{\boldmath{$\bar m$}}}

\newcommand{\be}{\begin{equation}}
\newcommand{\ee}{\end{equation}}

\newcommand{\beqn}{\begin{eqnarray}}
\newcommand{\eeqn}{\end{eqnarray}}
\newcommand{\AdS}{anti--de~Sitter }
\newcommand{\AAdS}{\mbox{(anti--)}de~Sitter }
\newcommand{\AAN}{\mbox{(anti--)}Nariai }
\newcommand{\AS}{Aichelburg-Sexl }
\newcommand{\pa}{\partial}
\newcommand{\pp}{{\it pp\,}-}
\newcommand{\ba}{\begin{array}}
\newcommand{\ea}{\end{array}}

\newcommand{\tr}{\textcolor{red}}
\newcommand{\tb}{\textcolor{blue}}
\newcommand{\tg}{\textcolor{green}}

\def\a{\alpha}
\def\g{\gamma}
\def\de{\delta}

\def\b{{\kappa_0}}

\def\E{{\cal E}}
\def\B{{\cal B}}
\def\R{{\cal R}}
\def\F{{\cal F}}

\def\e{e}
\def\bb{b}

\newtheorem{theorem}{Theorem}[section]

\title{{Static and radiating $p$-form black holes in the higher dimensional Robinson--Trautman class}}

\author{Marcello Ortaggio\thanks{ortaggio(at)math(dot)cas(dot)cz} \\
Institute of Mathematics, Academy of Sciences of the Czech Republic \\ \v Zitn\' a 25, 115 67 Prague 1, Czech Republic
 \\ \\
 Ji\v{r}\'{\i} Podolsk\'y\thanks{podolsky(at)mbox(dot)troja(dot)mff(dot)cuni(dot)cz} \ and Martin \v{Z}ofka\thanks{zofka(at)mbox(dot)troja(dot)mff(dot)cuni(dot)cz}
\\ Institute of Theoretical Physics, Faculty of Mathematics and Physics,\\
 Charles University in Prague, V Hole\v{s}ovi\v{c}k\'{a}ch 2, 180 00 Prague 8,  Czech Republic}

\maketitle

\abstract{We study Robinson--Trautman spacetimes in the presence of an aligned $p$-form Maxwell field and an arbitrary cosmological constant in ${n\ge 4}$ dimensions. As it turns out, the character of these exact solutions depends significantly on the (relative) value of $n$ and $p$. In {\em odd} dimensions the solutions reduce to static black holes dressed with an electric and a magnetic field, with an Einstein space horizon (further constrained by the Einstein--Maxwell equations) --- {both} the Weyl and Maxwell types are~D. {\em Even} dimensions, however, open up more possibilities. In particular, when ${2p=n}$ there exist non-static solutions describing black holes gaining (or losing) mass by receiving (or emitting) electromagnetic radiation. {In this case} the Weyl type is II (D) and the Maxwell type can be II (D) or N. Conditions under which the Maxwell field is self-dual (for odd $p$) are also discussed, and a few explicit examples presented. Finally, the case ${p=1}$ is special in all dimensions and leads to static metrics with a non-Einstein transverse space.}

\vspace{.2cm}

% 04.20.Jb  Exact solutions
% 04.50.+h  in more than four dimensions
% 04.40.Nr  Einstein-Maxwell spacetimes, spacetimes with fluids, radiation or classical fields

\tableofcontents

\section{Introduction}
\label{intro}

Gravity in more than four spacetime dimensions has attracted a lot of interest in recent years. In particular, several properties of higher dimensional black holes have been elucidated (see, e.g., the reviews \cite{EmpRea08,HolIsh12,Horowitzbook}). In addition to vacuum spacetimes, solutions of various theories with gauge fields have also been  investigated extensively (numerous references being given in \cite{EmpRea08,HolIsh12,Horowitzbook}). The simplest such theory one can consider is probably $n$-dimensional Einstein--Maxwell gravity in the presence of a single 2-form field $F_{\mu\nu}$. Among its solutions, an analog of the Reissner--Nordstr\"{o}m spacetime has been known for a long time \cite{Tangherlini63}, together with its generalizations admitting an Einstein horizon \cite{GibWil87} (in contrast, no $n>4$ exact solution is known that would extend the Kerr--Newman metric, except in the vacuum limit \cite{MyePer86}, or in other theories \cite{EmpRea08,HolIsh12,Horowitzbook}). However, since extended objects naturally couple to higher-rank forms, theories of gravity with $p>2$ forms are also of interest, especially from the standpoint of supergravity and string theory. A direct generalization of Einstein--Maxwell gravity is thus the $p$-form theory defined by
{
\be
  S=\frac{1}{16\pi}\int\d^n x\sqrt{-g}\left(R-2\Lambda-\frac{\b}{p}F^2\right) ,
\label{theory}
\ee
where $F^2=F_{\a_1\ldots\a_{p}}F^{\a_1\ldots\a_{p}}$ (and $\b$ is a constant taking into account different possible normalizations found in the literature, cf. also footnote~\ref{foot_dual})}, to which we shall restrict ourselves in the paper.

Although the theory~\eqref{theory} is in some respects very similar to standard Einstein--Maxwell gravity, the case $p=2$ possesses some distinct features. For example, it has been shown recently that asymptotically flat static black holes cannot couple to electric $p$-form fields when $(n+1)/2\le p\le n-1$ (and thus do not possess dipole hair) \cite{EmpOhaShi10} and that, for any $p>2$, static perturbations of the vacuum Schwarzschild--Tangherlini metric do not exist \cite{Guven89,ShiOhaTan11}. In addition, results of \cite{Durkeeetal10} indicate that electromagnetic radiation may have properties different from those of standard $n=4$, $p=2$ electrovac general relativity (except possibly for $2p=n$), as confirmed in \cite{Ortaggio14} in the case of test fields.

A relatively simple and yet rich class of exact solutions in $n=4$ general relativity is given by the Robinson--Trautman family \cite{RobTra62} (cf. the reviews \cite{Stephanibook,GriPodbook}), defined by the existence of a geodesic, shear-free, twist-free but expanding null vector field $\bk$. It includes static black holes with an arbitrary cosmological constant, their accelerating counterpart (the C-metric) and other spacetimes containing both gravitational and electromagnetic radiation, as well as pure radiation solutions such as the Vaidya metric. Some of these solutions have found useful applications also beyond general relativity, for example to describe $2+1$ black holes on a brane \cite{EmpHorMye00}, or in the context of the AdS/CFT correspondence \cite{BerRea14}.
Electrovac Robinson--Trautman spacetimes, in particular, have been investigated thoroughly in the presence of a Maxwell field $F_{\mu\nu}$ aligned with $\bk$ (already starting in \cite{RobTra62}) and can be of Petrov type II, D and III (not N and O), while the Maxwell field can be both of type D (``non-null'') and N (``null''), but it must be null if the Petrov type is III (see section~28.2 of \cite{Stephanibook} and section~19.6 of \cite{GriPodbook} for reviews and for a number of original references). Within this class, solutions of Petrov type D with a null Maxwell field \cite{RobTra62}\footnote{It is interesting to observe that the ``example'' given in section~7~(ii) of \cite{RobTra62} (eq.~(28.43) of \cite{Stephanibook})
is in fact (up to adding a cosmological constant) the {\em unique} \cite{VandenBergh89} solution of the 4D Einstein--Maxwell equations that is simultaneously both of Petrov type D and of Maxwell type N (in which case the Maxwell and Weyl tensor {\em necessarily} share a multiple principal null direction {by the Mariot--Robinson and the (generalized) Goldberg--Sachs theorems \cite{Stephanibook}, see also} \cite{CahSen67}) --- cf. also \cite{CahSen67,Leroy76,DebVanLer89} for various steps towards the complete proof \cite{VandenBergh89} of this statement.\label{footn_RTnull}} are of special interest, as they describe formation of black holes by gravitational collapse of purely electromagnetic radiation \cite{Senovilla14} (see \cite{Lemos98,Lemos99,DadFirMan12} for related earlier studies).

In \cite{PodOrt06} it was shown how the Robinson--Trautman line-element can be constructed in arbitrary dimension $n$ (see also \cite{Ortaggio07,OrtPodZof08,PraPraOrt07,SvaPod14,PodSva14} for additional results). In the context outlined above, it is thus of interest to study Robinson--Trautman solutions of the theory \eqref{theory} with arbitrary $n$ and $p$, which is the purpose of the present paper. We observe that the case $p=2$ has been already studied in detail in \cite{OrtPodZof08} (including a possible Chern-Simons term in odd dimensions). Similarly as in the vacuum case \cite{PodOrt06}, it turned out that the Robinson--Trautman class is much more restricted when $n>4$, and it essentially contains only static black hole spacetimes of Weyl type D (plus a few special non-static solutions \cite{PodOrt06,Ortaggio07,PraPraOrt07,OrtPodZof08}). However, the results of \cite{Durkeeetal10,Ortaggio14} mentioned above suggest that it need not be so for the theory~\eqref{theory}. As we will work out, this expectation turns out to be correct. In more detail, our results can be summarized as follows.

\begin{itemize}

\item In {\em odd} $n$ dimensions, Robinson--Trautman spacetimes coupled to a $p$-form Maxwell field with $2\le p\le n-2$ reduce to {\em static black holes} specified by four independent parameters related to mass, electric and magnetic field strengths, and cosmological constant (section~\ref{subsec_summ_gen}). These black holes have been already studied in \cite{BarCalCha12}, to which we add a few comments. Except for the case $p=2$ ($p=n-2$) of \cite{Tangherlini63}, they cannot be asymptotically flat (so there is no conflict with the results of \cite{EmpOhaShi10,Guven89,ShiOhaTan11}), but they can be asymptotically locally (A)dS in some cases.
The horizon is an Einstein space but must also obey further constraints following from the field equations. The metric is \eqref{ds_generic} with \eqref{H_generic} and the Maxwell field is \eqref{F_generic} (see the text for more details). Both the Weyl and  Maxwell tensors are of type D and share the same pair of doubly aligned (geodesic) null directions.

\item In {\em even} $n$ dimensions (with $2\le p\le n-2$), static black holes as those described above are also present (cf. also \cite{BarCalCha12}) and, again, the horizon geometry is constrained by the field equations (for example, for $p=2$ it must be almost-K\"ahler if the magnetic field is non-zero \cite{OrtPodZof08}). Moreover, in addition to static black holes,  there are also some exceptional solutions when $2p=n\pm2$ and $n\ge6$ (as first noticed in \cite{OrtPodZof08} for the case $n=6$, $p=2$), cf. section~\ref{subsubsec_summary_2p=n+2} for details. Generically, these exceptional metrics are non-static and of Weyl type II, while the Maxwell type is still D (but now possesses also a non-geodesic aligned null direction).

\item In even $n$ dimensions, the unique rank $p$ satisfying $2p=n$ (including, in particular, $n=4$ with $p=2$ \cite{RobTra62,Stephanibook,GriPodbook}) gives rise to an additional (and more interesting) new class of solutions (section~\ref{sec_2p=n}), consisting of the metric \eqref{ds_generic} with \eqref{grr_2p=n} and the Maxwell field \eqref{F_2p=n}. The Maxwell field is allowed to be of type II, D or N and in all these cases it can have a radiative term (thus giving concrete examples to the predictions of \cite{Durkeeetal10,Ortaggio14}), while the Weyl tensor can be of type II or D (more precisely, II(bd)/II(bcd) or D(bd)/D(bcd)) but with no radiative term for $n>4$. However, when the Maxwell type is N and $n>4$ then the Weyl type can only be D(bd)/D(bcd) (this is a significant difference w.r.t the $n=4$, $p=2$ case, which can be traced back to the absence of gravitational radiation in the vacuum higher dimensional Robinson--Trautman class \cite{PodOrt06}). Similarly as in \cite{Senovilla14}, some of these solutions can be used to describe {black hole formation (or white hole evaporation, by time-reversal) by collapse (emission) of electromagnetic radiation.} As in the static case, these black holes can be asymptotically locally (A)dS for certain choices of the parameters --- again, the horizon is an Einstein space and can be flat, in particular. For {\em odd} $p$, some of these solutions possess a {\em self-dual} $p$-form field, a property which is of interest in supergravity and string theory \cite{HenTei88}.

\item Both in odd and even dimensions, the rank $p=1$ (or its dual $p=n-1$) has special features and needs to be studied separately (appendix~\ref{app_limiting}). This results again in a family of static solutions of Weyl and Maxwell type D. In this case the transverse space cannot be an Einstein space since it ``feels'' the backreaction of the electromagnetic field (as opposed to the generic case $2\le p\le n-2$), which also defines a preferred direction here. Further, the transverse space  cannot be a space with an everywhere non-negative Ricci scalar. There exists no $n=4$ counterpart of these solutions.

\end{itemize}

The rest of the paper is organized as follows. In section~\ref{sec_geom} we describe our assumptions and set up the corresponding general form of the line-element (based on \cite{PodOrt06}) and of the Maxwell field. Section~\ref{Einstein} is devoted to a systematic  integration of the resulting Einstein--Maxwell equations --- it can well be skipped by readers not interested in those technicalities. Our results are summarized in section~\ref{sec_summary_gen} for the generic case $2p\neq n$, and in sections \ref{subsec_2p=n_summary}--\ref{subsec_null_Max} for the special case $2p=n$ (section~\ref{subsec_integr_2p=n} contains integration of a subset of the Einstein--Maxwell equations that are special when $2p=n$). Appendix~\ref{app_summaryRT}, largely based on \cite{PodOrt06,OrtPodZof08,SvaPod14,PodSva14}, summarizes certain general properties of Robinson--Trautman spacetimes useful in the paper, but also contains a few new observations (sections~\ref{subsubsec_weyl_III}--\ref{app_type_D(bd)}). Appendix~\ref{app_comments} discusses some geometrical properties of the transverse metric $h_{ij}$ of \eqref{ds_generic} (in particular, the black hole horizon) that follow from the Einstein equations (it partly overlaps with \cite{OrtPodZof08} when $p=2$ or $p=n-2$, and also summarizes certain observations of \cite{BarCalCha12}). Appendix~\ref{app_limiting} studies the special ranks $p=1$ and $p=n-1$, also showing that these are forbidden in the four dimensional Robinson--Trautman class.

\subsection*{Notation and conventions}

Throughout the paper we focus on $n>4$ dimensions, but large part of the results applies also to the $n=4$ case, on which we shall comment explicitly when important differences arise. We consider a $p$-form field $\bF=\frac{1}{p!}F_{\a_1\ldots\a_{p}}\d x^{\a_1}\wedge\ldots\wedge\d x^{\a_p}$ (with $1\le p\le n-1$)\footnote{The limiting value $p=n$ could also be included but, as well-known (cf., e.g., \cite{Duffvan80}), this is trivial in the sense that $\bF$ is simply given by the spacetime volume element (up to a constant rescaling) and acts on the geometry as an effective positive cosmological constant. Similarly, the dual case $p=0$ reduces to a constant scalar field.\label{footn_trivial}} that, {in the theory~\eqref{theory},} satisfies the source-free Maxwell equations {$\d^*\bF=0$, $\d\bF=0$ or, in components,}
\be
  (\sqrt{-g}\,F^{\mu\a_1\ldots\a_{p-1}})_{,\mu}=0 , \qquad F_{[\a_1\ldots\a_p,\mu]}=0 .
	\label{Maxwell}
\ee
The Maxwell field $\bF$ backreacts on the spacetime geometry via the energy-momentum tensor
\begin{equation}\label{Energy momentum}
  T_{\mu\nu} = \frac{\b}{8\pi} \left( F_{\mu\a_1\ldots\a_{p-1}} {F_\nu}^{\a_1\ldots\a_{p-1}}-\frac{1}{2p} g_{\mu\nu}F^2 \right) ,
\end{equation}
where $F^2=F_{\a_1\ldots\a_{p}}F^{\a_1\ldots\a_{p}}$. Note that $8\pi T=\b{(2p-n)}F^2/(2p)$. With \eqref{Energy momentum}, Einstein's equations with a cosmological constant $R_{\mu\nu} - \frac{1}{2} R g_{\mu\nu} + \Lambda g_{\mu\nu} = 8 \pi T_{\mu\nu}$ {(following from~\eqref{theory})} take the form
\begin{equation}\label{Ricci}
	R_{\mu\nu} = \frac{2}{n-2} \Lambda g_{\mu\nu} + \b\left[F_{\mu\a_1\ldots\a_{p-1}} {F_\nu}^{\a_1\ldots\a_{p-1}}-\frac{p-1}{p(n-2)} g_{\mu\nu}F^2 \right] .
\end{equation}

It may be useful to recall the well-known fact that for any solution of \eqref{Maxwell} and \eqref{Ricci} with a given $p$-form $\bF$, a ``discrete'' duality transformation $\bF\to {\mbox{\boldmath{$^*F$}}}$ gives rise to a dual solution\footnote{To be precise, in order for $T_{\mu\nu}$ to be invariant under $\bF\to {\mbox{\boldmath{$^*F$}}}$, $\b$ should be replaced by $\b/(p-1)!$ in \eqref{Energy momentum} and \eqref{Ricci} (which also shows how the limiting case $p=0$ can be formally incorporated in the discussion) --- this rescaling is not necessary in the case $2p=n$. While bearing this in mind, for compactness throughout the paper we will employ the simpler expressions \eqref{Energy momentum} and \eqref{Ricci}.\label{foot_dual}} where the same metric is coupled to the $(n-p)$-form ${\mbox{\boldmath{$^*F$}}}$, defined by $^*F_{b_1\ldots b_{n-p}}\equiv\frac{1}{p!}\epsilon^{a_1\ldots a_p}_{\qquad \ b_1\ldots b_{n-p}}F_{a_1\ldots a_p}$. In the special case $2p=n$, (anti-)self-dual $p$-form solutions ${\mbox{\boldmath{$^*F$}}}=\pm \bF$ (for which necessarily $\bF^2=0={\mbox{\boldmath{$^*F$}}}\cdot\bF$, where total contraction is understood) may exist if $p$ is {\em odd} (since $\bF$ is real and the signature Lorentzian, see, e.g., \cite{HenTei88}). Instead, for $2p=n$ with {\em even} $p$ (anti-)self-dual $p$-forms do not exist --- on the other hand, in that case there is a continuous $SO(2)$ duality symmetry (in addition to the discrete one mentioned above) which maps solutions into solutions (cf., e.g., \cite{Deseretal97} and appendix~A of \cite{Bremeretal98}).

\section{Robinson--Trautman geometry with aligned Maxwell fields}
\label{sec_geom}

We consider a $n$-dimensional spacetime that admits a non-twisting, non-shearing, expanding  geodesic null vector field~{\boldmath $k$}. The associated Robinson--Trautman line-element was obtained in adapted coordinates in \cite{PodOrt06}. The corresponding curvature has been fully computed recently in \cite{PodSva14}, showing in particular that the spacetime is generically of aligned Weyl type I(b).

It is the purpose of this paper to determine Robinson--Trautman spacetimes in the presence of a Maxwell $p$-form field {in the theory~\eqref{theory},} i.e., such that the Einstein equations \eqref{Ricci} are satisfied, along with the Maxwell equations \eqref{Maxwell}. However, we shall restrict to the case of Maxwell fields that are {\em aligned} with {\boldmath $k$} (so that $\bF$ is of type II or more special).\footnote{We refer to the boost weight (b.w.) classification of a general tensor \cite{Milsonetal05} --- cf., e.g., \cite{BerSen01,Coleyetal04vsi,Milson04,HerOrtWyl13} for further results in the particular case of 2-forms. Let us observe that since {\boldmath $k$} is shearfree and expanding here, a result of \cite{Durkeeetal10} implies that $\bF$ cannot be doubly aligned with it (i.e., it cannot be of type N), except possibly when $2p=n$ in even dimensions. We will prove that type N fields do indeed exist in that special case (section~\ref{sec_2p=n}).\label{footn_null}} This means that the components of $\bF$ of b.w.  +1 vanish, which by \eqref{Ricci} implies that the Ricci tensor components of b.w.  +2 and +1 must also vanish, i.e., the Ricci type is II (or more special) aligned with $\bk$. With this condition, we can take advantage of previous results of \cite{PodOrt06} (summarized in theorem~\ref{theor_RT} of appendix~\ref{app_summaryRT}) asserting that the spacetimes in question can be represented by the line-element
\beqn
  & & \d s^2=r^{2}h_{ij}\left(\d x^i+ W^{i}\d u\right)\left(\d x^j+ W^{j}\d u\right){-}2\,\d u\d r-2H\d u^2 , \nonumber \\
	& & h_{ij}=h_{ij}(u,x) , \qquad W^{i}={\alpha}^i(u,x)+r^{1-n}{\beta}^i(u,x) \label{geo_metric_text} .
\eeqn
Here the adapted coordinates $(u,r,x^1,\ldots,x^{n-2})$ are used (Latin indices $i,j,\ldots$ or $i_1, i_2,\dots$ etc. range over $1,\ldots,n-2$ and label the spatial coordinates $x^i$, sometimes collectively denoted simply as $x$) such that
\be
 \bk\equiv k^\mu\pa_\mu=\pa_r , \qquad {k_\mu\d x^\mu={-}\d u } ,
\label{k}
\ee
where $r$ is an affine parameter along the generator $\bk$ of the null hypersurfaces ${u=\hbox{const.}}$ In such coordinates, the assumed alignment condition on $\bF$ takes the form
\be
	F_{r i_1\ldots i_{p-1}}=0 ,
	\label{F_aligned}
\ee
or, equivalently, $F^{u i_1\ldots i_{p-1}}=0$, while the alignment conditions on the Ricci tensor (automatically satisfied by \eqref{geo_metric_text}) read $R_{rr}=0=R_{ri}$. By construction, the corresponding components of the Einstein equation \eqref{Ricci} are thus identically satisfied (note also that \eqref{F_aligned} implies $T_{rr} =0=T_{ri}$, cf. \eqref{Energy momentum}).

In the rest of the paper we thus need to study only the remaining Einstein equations for $R_{ij}$, $R_{ur}$, $R_{ui}$ and $R_{uu}$. These will contain terms depending on the Maxwell field. Hence, in order to proceed it will be convenient to first fix the $r$-dependence of $\bF$ using Maxwell's equations \eqref{Maxwell}.

For later calculations it is also useful to note that
\begin{equation}\label{determinants}
  \sqrt{-g} = r^{n-2}\,\sqrt h ,
\end{equation}
where $g\equiv \det g_{\mu\nu}$ and $h(u,x) \equiv \det h_{ij}$.

\section{Integration of the Einstein--Maxwell field equations}
\label{Einstein}

\subsection{Maxwell equations, step one}

\label{subsec_Max1}

With \eqref{F_aligned}, the ``geometrical'' equations $F_{[\a_1\ldots\a_p,\mu]}=0$ give
\begin{eqnarray}
 & & F_{i_1\ldots i_p,r} = 0, \label{F_ijr} \\
 & & F_{[i_1\ldots i_p,j]}= 0, \label{F_ijk} \\
 & & F_{uji_1\ldots i_{p-2},r} = (p-1)F_{ur[i_1\ldots i_{p-2},j]} , \label{F_uir} \\
 & & F_{ji_1\ldots i_{p-1},u} = pF_{u[i_1\ldots i_{p-1},j]} . \label{F_iju}
\end{eqnarray}
Using relation (\ref{determinants}), the ``dynamical'' equations $(\sqrt{-g}\,F^{\mu u r i_1 \ldots i_{p-3}})_{,\mu}=0$ (not present for {$p\le2$}) and $(\sqrt{-g}\, F^{\mu u i_1 \ldots i_{p-2}})_{,\mu}=0$ (not present for $p\le1$) read, respectively,
\be
  (\sqrt{h}\,F^{urji_1\ldots i_{p-3}})_{,j}=0 , \qquad (r^{n-2}\,F^{uri_1\ldots i_{p-2}})_{,r}=0 . \label{dynamical_1}
\ee
As it turns out, the remaining equations $(\sqrt{-g}\,F^{\mu i_1 \ldots i_{p-1}})_{,\mu}=0$ and $(\sqrt{-g}\,F^{\mu r i_1 \ldots i_{p-2}})_{,\mu}=0$ become significantly simpler once some of the Einstein equations are enforced, and it is thus convenient to postpone their discussion to section~\ref{subsec_Max2}.

The $r$-dependence of the Maxwell field is thus completely determined by \eqref{F_ijr}, the second of \eqref{dynamical_1} and \eqref{F_uir}, and can be summarized as
\beqn
 & & F_{i_1\ldots i_p}= \bb_{i_1\ldots i_p} ,  \label{Max0_r} \\
 & & F_{uri_1\ldots i_{p-2}}=r^{2p-2-n}\,\e_{i_1\ldots i_{p-2}} , \label{Max0_ur} \\
 & & F_{uji_1\ldots i_{p-2}}=r^{2p-1-n}\frac{p-1}{2p-1-n}\,\e_{[i_1\ldots i_{p-2},j]}+{f}_{ji_1\ldots i_{p-2}} \qquad (2p\neq n+1), \label{Max-1_r}
\eeqn
where lowercase symbols $\bb$, $\e$ or $f$ denote integration functions independent of $r$. {Since we are restricting to $1\le p\le n-1$, $\bb_{i_1\ldots i_p}=\bb_{[i_1\ldots i_p]}$ can be non-zero only for $1\le p\le n-2$, and $\e_{i_1\ldots i_{p-2}}=\e_{[i_1\ldots i_{p-2}]}$ for $2\le p\le n-1$ (for $p=2$ the term $\e_{i_1\ldots i_{p-2}}$ obviously reduces to a scalar function $e$).} From now on we will use the convention that indices of $\bb$, $\e$ and $f$ are raised with the spatial metric $h^{ij}$ so that, e.g.,
\beqn
	& & \bb^{i_1\ldots i_{p}}= h^{i_1j_1}\ldots h^{i_{p}j_{p}}\,\bb_{j_1\ldots j_{p}}	, \qquad \e^{i_1\ldots i_{p-2}}= h^{i_1j_1}\ldots h^{i_{p-2}j_{p-2}}\,\e_{j_1\ldots j_{p-2}} , \nonumber \\
	& & f^{i_1\ldots i_{p-1}}= h^{i_1j_1}\ldots h^{i_{p-1}j_{p-1}}\,f_{j_1\ldots j_{p-1}} .
\eeqn

In the special case with $2p=n+1$ ($n\ge 5$ odd, $p\ge3$), eq.~\eqref{Max-1_r} is replaced by\footnote{Obviously, for dimensional reasons we should write $\ln (r/r_0)$ instead of $\ln r$, where $r_0$ is an $r$-independent coefficient with the dimension of length. However, $r_0$ can be absorbed in the following term ${f}_{ji_1\ldots i_{p-2}}$ and thus for brevity we will omit it. Similar comments apply to further logarithmic terms that will appear in other expressions in the following, and will not be repeated there.}
\be
 F_{uji_1\ldots i_{p-2}}={\frac{n-1}{2}}\e_{[i_1\ldots i_{p-2},j]}\ln r+{f}_{ji_1\ldots i_{p-2}} \qquad (2p=n+1). \label{Max-1_r_spec}
\ee
(However, we will see in section~\ref{subsec_Max2} that the above logarithmic term must in fact vanish.) From \eqref{Max0_r} we observe, in particular, that the magnetic components $F_{i_1\ldots i_p}$ are always $r$-independent, whereas the electric components $F_{uri_1\ldots i_{p-2}}$ become $r$-independent only in the special case $2p=2+n$ ($n$ even).

Using \eqref{Max0_ur} and \eqref{Max0_r}, the first of \eqref{dynamical_1} and \eqref{F_ijk} can be rewritten simply as
\be
	(\sqrt{h}\,\e_{}^{ji_1\ldots i_{p-3}})_{,j}=0 \quad (p\ge 3) , \qquad {\bb_{[i_1\ldots i_p,j]}= 0 } . \label{diverg_elec}
\ee

For later purposes, it will be useful to define the following $r$-independent quantities built out of the electric and magnetic parts of the Maxwell field (which will enter some of the Einstein equations)
\beqn
 & & \E_{ij}^2\equiv \e_{ik_1\ldots k_{p-3}}\,\e_{j}^{\ k_1\ldots k_{p-3}}  , \qquad \E^2\equiv h^{ij}\,\E_{ij}^2 \qquad (p\ge 3) , \label{E} \\
 & & \B_{ij}^2\equiv \bb_{ik_1\ldots k_{p-1}}\,\bb_{j}^{\ k_1\ldots k_{p-1}} , \qquad \B^2\equiv h^{ij}\,\B_{ij}^2 \qquad (p\le n-2) . \label{B}
\eeqn
Clearly $\E_{ij}^2=\E_{(ij)}^2$ and $\B_{ij}^2=\B_{(ij)}^2$. In the case $p=2$ the indices $i,j$ disappear from $\e_{i_1\ldots i_{p-2}}$ and $\E_{ij}^2$ and we simply have $\E=\e$, while $\B_{ij}^2$ is identically zero for $p>n-2$.  Obviously $\E^2=0\Leftrightarrow \e_{i_1\ldots i_{p-2}}=0$ and $\B^2=0\Leftrightarrow \bb_{i_1\ldots i_p}=0$ since $h_{ij}$ is positive definite.

In particular, the invariant $F^2$ (useful in the following) can be written as
\be
 F^2=-r^{2(p-n)}p(p-1)\E^2+r^{-2p}\B^2 .
 \label{F_invariant}
\ee

\subsection{Einstein equations for $R_{ij}$ and $R_{ur}$}

\label{subsec_Rur}

Knowing the $r$-dependence of the Maxwell field, we can now consider the remaining Einstein equations. It is convenient to start from the equations for the Ricci components $R_{ij}$ and $R_{ur}$ (those of the highest remaining b.w., namely zero). From~\eqref{Ricci} with \eqref{geo_metric_text}, \eqref{Max0_r}, {\eqref{Max0_ur},} \eqref{F_invariant} (using the definitions \eqref{E}, \eqref{B}), these read
\beqn
	& & R_{ij} = \frac{2}{n-2} \Lambda r^2h_{ij} + \b\left\{r^{2(p+1-n)}(p-1)\left[-(p-2)\E^2_{ij}+\frac{p-1}{n-2}\E^2h_{ij}\right] \nonumber \right. \\
	& & \qquad\qquad\qquad\qquad\qquad\qquad\qquad \left. {}+r^{2(1-p)}\left[\B^2_{ij}-\frac{p-1}{p(n-2)}\B^2h_{ij}\right]\right\} , \label{Rij} \\
	& & {-} R_{ur} = \frac{2}{n-2} \Lambda - \b\frac{p-1}{n-2}\left[r^{2(p-n)}\E^2(n-p-1)+r^{-2p}\B^2p^{-1}\right] . \label{Rur}
\eeqn

The component $R_{ij}$ of the metric \eqref{geo_metric_text} is reproduced in appendix~\ref{app_summaryRT} as eq.~\eqref{Rij_general}. Comparing this with \eqref{Rij} immediately reveals that ${\beta}^i=0$ (this follows by comparing various powers of $r$ in \eqref{Rij} and \eqref{Rij_general}, which implies that ${\beta}^i{\beta}^j$ is either zero or proportional to $h^{ij}$ --- the latter option is however impossible, cf. also \cite{PodOrt06}).\footnote{One might think that the (dual) cases $p=1$ and $p=n-1$ escape this conclusion since the $r^{2(2-n)}$ term of \eqref{Rij_general} falls off in those cases as the term $r^{2(p+1-n)}$ or $r^{2(1-p)}$ of \eqref{Rij}, respectively. However, for $p=1$ [$p=n-1$] the $r^{2(p+1-n)}$ [$r^{2(1-p)}$] term of \eqref{Rij} vanishes identically, cf.~\eqref{E} and \eqref{B}, so that the conclusion ${\beta}^i=0$ remains true.}
One can further perform a coordinate transformation (at least locally) to set ${\alpha}^i=0$ \cite{PodOrt06}. From now on we shall thus have in \eqref{geo_metric_text}
\be
 W^i=0 , \label{f=0=e}
\ee
and therefore $g^{ri}=0=g_{ui}$, which will simplify several expressions. In particular, the Weyl type will thus be {II(d)} or more special, aligned with $\bk$ (cf. theorem~\ref{theor_RT}).

Using \eqref{f=0=e}, the component \eqref{Rij_general} now simplifies drastically {(cf.~\eqref{Rij_general_W=0})}. By comparing its various powers of $r$ with those of \eqref{Rij}, after $r$-integration (and contraction with $h^{ij}$ when necessary) one readily determines the $r$-dependence of the metric function $H$

\begin{eqnarray}
	& & 2H=\frac{{\cal R}}{(n-2)(n-3)}{+}\frac{2(\ln\sqrt{h})_{,u}}{n-2}\,r-\frac{2\Lambda}{(n-1)(n-2)}\,r^2-\frac{\mu}{r^{n-3}} \nonumber\\
	& & \qquad\qquad {}+\frac{\b}{n-2}\left[\frac{p-1}{n+1-2p}\frac{\E^2}{r^{2(n-p-1)}}-\frac{1}{p(n-1-2p)}\frac{\B^2}{r^{2(p-1)}}\right] \qquad (2p\neq n\pm1), \label{grr_final}
\end{eqnarray}
where $\mu$ is an arbitrary integration function independent of $r$, and $\R$ is the Ricci scalar associated with the spatial metric $h_{ij}$.

In the special odd dimensional cases $2p=n\pm1$ the last two terms of \eqref{grr_final} should be replaced, respectively, by
\begin{eqnarray}
	& & +\frac{\b}{n-2}\left[-\frac{n-1}{2}\E^2\frac{\ln r}{r^{n-3}}+\frac{1}{n+1}\frac{\B^2}{r^{n-1}}\right] \qquad (2p=n+1) , \label{H_log+} \\
	& & +\frac{\b}{n-2}\left[\frac{n-3}{4}\frac{\E^2}{r^{n-1}}-\frac{2}{n-1}\B^2\frac{\ln r}{r^{n-3}}\right] \ \ \qquad (2p=n-1) . \label{H_log-}
\end{eqnarray}

In addition, as a further consequence of the Einstein equation for $R_{ij}$, the following constraints (coming from terms of order $r^0$, $r$, $r^{2(p+1-n)}$ and $r^{2(1-p)}$, respectively) must be satisfied (also when $2p=n\pm1$)
\beqn
 & & \R_{ij}=\frac{\R}{n-2}h_{ij}  \qquad {(p\neq 1, n-1)} , \label{constrijr} \\
 & & h_{ij,u} =\frac{2(\ln\sqrt{h})_{,u}}{n-2}h_{ij} , \label{constrijs} \\
 & & \E^2_{ij}=\frac{\E^2}{n-2}h_{ij} \qquad {({3\le p\le n-2}, \ 2p\neq n)} , \label{eqE} \\
 & & \B^2_{ij}=\frac{\B^2}{n-2}h_{ij} \qquad {({2\le}p\le n-2, \ 2p\neq n)}, \label{eqB}
\eeqn
where  ${\R_{ij}}$ is the Ricci tensor associated with $h_{ij}$ (so that $\R=h^{ij}\R_{ij}$), and the general identity $h^{ij}h_{ij,u}=2(\ln \sqrt{h})_{,u}$ has been used. For $p=2$ eq.~\eqref{eqE} becomes an identity. For $2p=n$ eqs.~\eqref{eqE} and \eqref{eqB} are replaced by the following single equation
\be
 \frac{1}{4}(n-2)(n-4)\left(\E^2_{ij}-\frac{\E^2}{n-2}h_{ij}\right)=\B^2_{ij}-\frac{\B^2}{n-2}h_{ij} \qquad (2p=n). \label{eqEB}
\ee
Eq.~\eqref{eqEB} is satisfied identically in the case $n=4$, $p=2$. Note that \eqref{eqE}, \eqref{eqB} and \eqref{eqEB} imply with \eqref{Rij} that in all cases $R_{ij}\propto h_{ij}$ (as can also be seen from {\eqref{Rij_general_W=0} with  \eqref{constrijr}).}

As indicated above, equations \eqref{constrijr}, \eqref{eqE} and \eqref{eqB} do not hold in the limiting dual cases $p=1$ (or $p=n-1$) --- these are special since the $\B^2_{ij}$ (or $\E^2_{ij}$) terms behave as $r^0$ in \eqref{Rij}, \eqref{Rur} (and therefore in \eqref{grr_final}), and thus effectively act as sources for the transverse geometry. Since further differences will also arise in the remaining Einstein equations, we present all the results for $p=1, n-1$ in appendix~\ref{app_limiting} and from now on we assume $p\neq1, n-1$.

As in \cite{PodOrt06}, relation (\ref{constrijr}) means that, at any given $u=u_0=\,$const, the spatial metric $h_{ij}(x,u_0)$ must describe a $(n-2)$-dimensional Riemannian Einstein space.
It is well-known {(see, e.g., p.76 of \cite{Petrov})} that for $n>4$ (i.e., $n-2>2$) this implies that ${\mathcal R}_{,i}=0$, so that ${\mathcal R}$ can depend only on the coordinate $u$ (additionally, for $n=5$ the metric $h_{ij}$ must be of constant curvature since it is 3-dimensional and Einstein). For $n=4$ eq.~(\ref{constrijr})  is instead an identity.

Equation (\ref{constrijs}) gives \cite{PodOrt06}
\be
 h_{ij}=h^{1/(n-2)}\,\gamma_{ij}(x)  \qquad \mbox{ where } \det \gamma_{ij}=1 , \label{h_u}
\ee
so that $h_{ij}$ can depend on $u$ only via the conformal factor $h^{1/(n-2)}$.

Eqs.~\eqref{eqE} and \eqref{eqB} (or \eqref{eqEB}) constrain both the Maxwell field and the metric $h_{ij}$, and the permitted relation between $p$ and $n$ --- some comments are given in appendix~\ref{app_comments} (see also \cite{BarCalCha12}, and \cite{OrtPodZof08} for the case $p=2$).

The component $R_{ur}$ of the metric \eqref{geo_metric_text} with \eqref{f=0=e} is reproduced in appendix~\ref{app_summaryRT} as eq.~\eqref{Rur_general}. Substituting \eqref{grr_final} into \eqref{Rur_general} and comparing with \eqref{Rur}, one finds that the corresponding Einstein equation is satisfied identically (including the cases $2p=n\pm1$). We finally observe that \eqref{constrijr} (with \eqref{f=0=e}) further restricts the Weyl tensor to be of type II(bd), aligned with $\bk$ (see theorem~\ref{theor_RT}).

\subsection{Maxwell equations, step two ($2p\neq n$)}

\label{subsec_Max2}

We can now turn to the remaining {set of the} Maxwell equations, namely  $(\sqrt{-g}\, F^{\mu i_1 \ldots i_{p-1}})_{,\mu}=0$ and $(\sqrt{-g}\, F^{\mu r i_1 \ldots i_{p-2}})_{,\mu}=0$, which were not considered in section~\ref{subsec_Max1}. As it turns out, from now on it will be necessary to consider the case $2p=n$ separately. Hereafter we thus restrict to the ``generic'' case $2p\neq n$, while the corresponding analysis for the special case $2p=n$ will be given later on in section~\ref{sec_2p=n}.

Using \eqref{Max0_r}, \eqref{Max-1_r}, and \eqref{f=0=e}, $(\sqrt{-g}\, F^{\mu i_1 \ldots i_{p-1}})_{,\mu}=0$ contains three different powers of $r$, i.e., $r^{n-2p-2}$, $r^{n-2p-1}$, $r^{-2}$, generically leading, respectively, to
\begin{equation}
    (\sqrt{h}\, \bb^{ji_1 \ldots i_{p-1}})_{,j}= 0 , \qquad  {f}_{ j_1 \ldots j_{p-1}} = 0 , \qquad \e_{[j_2 \ldots j_{p-1}, j_1]} = 0  . \label{diverg_magn}
\end{equation}

Note that the last two equations mean  (see \eqref{Max-1_r})
\be
  F_{u j_1 \ldots j_{p-1}} = 0   .
  \label{Fuigeneric}
\ee
{With \eqref{F_aligned}} this implies that $\bF$ is aligned also with the null vector $\bl={-}\pa_u+H\pa_r$ (and is thus of type D), and will be important in the following (as a consequence, $\bF$ cannot be of type N, as found in \cite{Durkeeetal10}). Eq.~\eqref{F_iju} thus becomes
\be
 \bb_{i_1 \ldots i_{p},u}=0   . \label{b_u}
\ee

Next, from $(\sqrt{-g}\, F^{\mu r i_1 \ldots i_{p-2}})_{,\mu}=0$ one gets {(using~\eqref{Fuigeneric})}
\begin{equation}
    (\sqrt{h}\, \e^{ i_1 \ldots i_{p-2}})_{,u} =0  ,
\end{equation}
which with \eqref{h_u} simply gives
\begin{equation}
    \e_{ i_1 \ldots i_{p-2}}=h^{\frac{2p-n-2}{2(n-2)}}\,\tilde\e_{i_1 \ldots i_{p-2}}(x)  . \label{e_u}
\end{equation}

Note that the above results apply also in the {$2p=n+1$} case \eqref{Max-1_r_spec} with a logarithmic term, which thus in fact vanishes. For later purposes it is useful to observe that \eqref{e_u} and \eqref{b_u} imply, respectively (recall \eqref{E} and \eqref{B}),
$\E^2_{ij}=h^{(p-n+1)/(n-2)}\tilde\E^2_{ij}(x)$ and $\B^2_{ij}=h^{(1-p)/(n-2)}\tilde\B^2_{ij}(x)$ (of course $\tilde\E^2_{ij}$ ({or} $\tilde\B^2_{ij}$) can be expressed in terms of {$\tilde\e_{i_1 \ldots i_{p-2}}$ ({or} $\tilde\bb_{i_1 \ldots i_{p}}$)} and $\gamma^{ij}(x)$, if desired), and thus $\E^2=h^{(p-n)/(n-2)}\tilde\E^2(x)$, $\B^2=h^{-p/(n-2)}\tilde\B^2(x)$. These in turn imply

\be
 (n-2)(\E^2)_{,u}=-2(n-p)\E^2(\ln\sqrt{h})_{,u}  , \qquad (n-2)(\B^2)_{,u}=-2p\B^2(\ln\sqrt{h})_{,u} . \label{EB_u}
\ee
which will be useful later on.

We observe that above we have not employed eqs.~\eqref{constrijr}--\eqref{eqB}, so that the results of the present section~\ref{subsec_Max2} apply also to the cases $p=1,n-1$.

\subsection{Einstein equation for $R_{ui}$ ($2p\neq n$; $p\neq 1,n-1$)}

\label{subsec_Rui}

As remarked above, we now consider $2p\neq n$. Recalling \eqref{f=0=e} we have ${g_{ui}=0=g^{ri}}$. Using \eqref{F_aligned} and \eqref{Fuigeneric}, we also obtain ${F_{u\a_1\ldots\a_{p-1}} {F_i}^{\a_1\ldots\a_{p-1}}=0}$ (and thus $T_{ui}=0$). The corresponding Einstein equation~\eqref{Ricci} thus simplifies considerably to
\be
   R_{ui} = 0 ,
\ee
where the explicit Ricci component is given by \eqref{Rui_general}. Employing \eqref{grr_final} shows that this equation contains distinct powers of $r$, namely $r^0$, $r^{-1}$, $r^{2-n}$, $r^{2p-2n+1}$, $r^{1-2p}$. The term of order $r^0$ vanishes identically thanks to \eqref{constrijs}, while the remaining terms immediately give, respectively, the following conditions
\beqn
 (n-4)\, \R_{,i}=0 & & \qquad {(p\neq 1, n-1)} ,  \label{constrijri} \\
   \mu_{,i} =0, & &  \label{constrijsi} \\
 (2p-n-2) (\E^2)_{,i}=0 \  & & \qquad (p\ge 2) , \label{eqEi} \\
 (2p-n+2) (\B^2)_{,i}=0.& &  \label{eqBi}
\eeqn
Eq.~\eqref{constrijri} is an identity due to \eqref{constrijr} (as mentioned in section~\ref{subsec_Rur}). One arrives at \eqref{constrijri}--\eqref{eqBi} also for $2p=n\pm1$ after replacing \eqref{grr_final} by the corresponding form of $H$ containing the logarithmic terms ({eqs.~\eqref{H_log+}, \eqref{H_log-}}). Thus, generically, the Ricci curvature $\R$ of the transverse ${(n-2)}$-dimensional Riemannian space, the ``mass'' parameter $\mu$, the electric scalar $\E^2$ and the magnetic scalar $\B^2$ must all be \emph{independent of the spatial coordinates}. However, $\E^2$ and $\B^2$ can depend on $x$ in the special cases ${2p=n+2}$ and ${2p=n-2}$, respectively ($n$ necessarily even) --- this will have some consequences in section~\ref{subsec_Ruu}. The case ${n=4}$ is also special in that $\R$ can depend on the coordinates $x$ --- however, since the case ${n=4}$, $p=2$ is already well-known \cite{RobTra62,Stephanibook,GriPodbook}, only the cases $n=4$, $p=1,3$ remain to be studied. These are precisely the ones dealt with in appendix~\ref{app_limiting}, so that from now on we can restrict in the main text to $n>4$ with no loss of generality.

\subsection{Einstein equation for $R_{uu}$ ($2p\neq n$; $p\neq 1,n-1$; $n>4$)}

\label{subsec_Ruu}

\subsubsection{Case $2p\neq n,  n\pm 1, n\pm 2$}

\label{subsubsec_Ruu_gener}

Using $g_{uu}=-2H$, in view of \eqref{Fuigeneric} we obtain ${F_{u\a_1\ldots\a_{p-1}} {F_u}^{\a_1\ldots\a_{p-1}}=
2H(p-1)\,r^{2(p-n)}\E^2 }$. With \eqref{F_invariant}, the last Einstein equation \eqref{Ricci} can thus be written in the simple form ${R_{uu}= 2HR_{ur}}$, where $R_{ur}$ is the ``on-shell'' Ricci component \eqref{Rur}. Using \eqref{Ruu_general} this means
\be
 {-}r^2(r^{-2}H)_{,r}(\ln\sqrt{h})_{,u}{+}(n-2)r^{-1}H_{,u}+r^{-2}\triangle H -(\ln\sqrt{h})_{,uu}-\frac{1}{4}h^{il}h^{jk}\,h_{ij,u}h_{kl,u}=0 ,  \label{Ruu_eq}
\ee
where $\triangle$ is the Laplace operator associated with $h_{ij}$ (cf. appendix~\ref{subapp_ricci}). Substituting~\eqref{grr_final} one easily sees that (for $2p\neq n\pm1, n\pm2$) this equation contains terms proportional to $r^0$, $r^{-1}$, $r^{2-n}$, $r^{2p-2n+1}$,  $r^{1-2p}$. Terms of order $r^0$ and $r^{-1}$ do not contain the Maxwell field and, as shown in the vacuum case \cite{PodOrt06}, vanish identically after using \eqref{constrijs} and certain identities (see also the appendix of \cite{SvaPod14}). Terms of order $r^{2p-2n+1}$ and $r^{1-2p}$ vanish thanks to \eqref{EB_u}. The only surviving $r^{2-n}$ term fixes the $u$-dependence of $\mu(u)$ (as in vacuum)
\be
 (n-2)\mu_{,u}=-(n-1)\mu(\ln\sqrt{h})_{,u} \qquad (2p\neq n\pm 2, n\pm 1, n>4). \label{Ruu_generic}
\ee

As in \cite{OrtPodZof08}, taking the $\pa_i$ derivative of \eqref{Ruu_generic} (with \eqref{constrijsi}) and of \eqref{EB_u} (with \eqref{eqEi} and \eqref{eqBi}) gives immediately that $(\ln\sqrt{h})_{,ui}=0$ (unless $\mu=\E^2=\B^2=0$, which yields $\bF=0$ and thus a vacuum spacetime \cite{PodOrt06}). This means $h(u,x)=U(u)X(x)$, but one can set $U(u)=1$ by rescaling the coordinates $u$ and $r$ (which preserves the line element, see section~4 of \cite{PodOrt06} and section~4.1 of \cite{OrtPodZof08} for details). Hence, without loss of generality, from now on we can restrict to the case $h_{,u}=0$. With \eqref{Ruu_generic}, \eqref{constrijsi}, \eqref{constrijri}, \eqref{EB_u}, \eqref{eqEi}, \eqref{eqBi}  this implies
\be
 h_{ij}=h_{ij}(x) , \qquad \R=\mbox{const} , \qquad \mu=\mbox{const} , \qquad \E^2=\mbox{const} , \qquad \B^2=\mbox{const} \qquad (n>4) , \label{constants_gen}
\ee
so that $H$ is a function of $r$ only (cf.~\eqref{grr_final}) and \eqref{Ruu_generic} is satisfied identically. Up to a further (constant) coordinate rescaling, one can fix the normalization of $\R$ such that $\R=0,\pm(n-2)(n-3)$ \cite{PodOrt06,OrtPodZof08}. We also observe that \eqref{e_u} now implies $ \e_{ i_1 \ldots i_{p-2}}=\e_{ i_1 \ldots i_{p-2}}(x)$.

\subsubsection{Case $2p=n\pm1$}

For $2p=n\pm1$ ($n$ odd) there are additional terms of order $r^{2-n}$ in \eqref{Ruu_eq} (due to the logarithmic terms in $H$, cf.~{\eqref{H_log+}, \eqref{H_log-}}) and \eqref{Ruu_generic} is replaced by
\beqn
 & & \mu_{,u}=-\frac{n-1}{n-2}(\ln\sqrt{h})_{,u}\left[\mu-\frac{\b}{2(n-2)}\E^2\right]   \quad\qquad\qquad (2p=n+1) , \label{Ruu_2p=n+1} \\
 & & \mu_{,u}=-\frac{n-1}{n-2}(\ln\sqrt{h})_{,u}\left[\mu-\frac{2\b}{(n-1)^2(n-2)}\B^2\right]  \qquad (2p=n-1) . \label{Ruu_2p=n-1}
\eeqn

However, here we still have $(\E^2)_{,i}=0=(\B^2)_{,i}$ (cf. \eqref{eqEi}, \eqref{eqBi}), so that as in section~\ref{subsubsec_Ruu_gener} we can choose coordinates such that \eqref{constants_gen} holds (so that \eqref{Ruu_2p=n+1} and \eqref{Ruu_2p=n-1} are identically satisfied). In the end, the only difference in this case is thus the presence of the logarithmic terms in $H$.

\subsubsection{Case $2p=n\pm2$}

\label{subsubsec_Ruu_2p=n+2}

Also for $2p=n\pm2$ ($n$ even) there are additional terms of order $r^{2-n}$ in \eqref{Ruu_eq} (now due to \eqref{eqEi}, \eqref{eqBi}) so that instead of \eqref{Ruu_generic} we obtain
\beqn
 & & \mu_{,u}=-\frac{n-1}{n-2}\mu(\ln\sqrt{h})_{,u}{-}\frac{\b n}{(n-2)^2}\triangle (\E^2)  \quad\qquad (n\ge 6, 2p=n+2) , \label{Ruu_2p=n+2} \\
 & & \mu_{,u}=-\frac{n-1}{n-2}\mu(\ln\sqrt{h})_{,u}{-}\frac{2\b}{(n-2)^3}\triangle (\B^2) \qquad (n\ge 6, 2p=n-2) \label{Ruu_2p=n-2} .
\eeqn

Recall that in the above two cases we generically have (eqs. \eqref{eqEi} and \eqref{eqBi}) $(\B^2)_{,i}=0\neq(\E^2)_{,i}$ (for $2p=n+2$)  and $(\E^2)_{,i}=0\neq(\B^2)_{,i}$ (for $2p=n-2$). The argument of section~\ref{subsubsec_Ruu_gener} leading to $h(u,x)=U(u)X(x)$ (and then to $h_{ij}=h_{ij}(x)$)  will thus still work in both cases provided $\B^2\neq0$ (for $2p=n+2$) or $\E^2\neq0$ (for $2p=n-2$). When these conditions are met, in suitable coordinates we thus arrive at (cf. a more detailed discussion in section~4.3 of \cite{OrtPodZof08} for the particular case $n=6$, $p=2$)
\beqn
 & & h_{ij}=h_{ij}(x) , \quad \R=\mbox{const} , \quad \mu_{,u}={-}{\frac{\b n}{(n-2)^2}}\triangle (\E^2) , \quad \B^2=\mbox{const}\neq0 \quad (2p=n+2) , \label{Ruu_2p=n+2_spec} \\
 & & h_{ij}=h_{ij}(x) , \quad \R=\mbox{const} , \quad \E^2=\mbox{const}\neq0 , \quad \mu_{,u}={-}{\frac{2\b}{(n-2)^3}}\triangle (\B^2) \quad (2p=n-2) . \label{Ruu_2p=n-2_spec}
\eeqn
In both cases we have $\mu_{,i}=0$ and $(\E^2)_{,u}=0=(\B^2)_{,u}$ {(see \eqref{EB_u})}, {so that $\pa_u$ is a Killing vector (twisting iff $H_{,i}\neq 0$).} With the above equations this implies that $\mu_{,u}$ and $\triangle (\E^2)$ (respectively, $\triangle (\B^2)$) are both constants. If the transverse space with metric $h_{ij}$ is assumed to be {{\em compact} (as for black hole spacetimes), then it follows (see, e.g., \cite{Bochner48,YanBocbook,KobNom2}) that $\E^2$ (respectively, $\B^2$) and thus $\mu$ are constant, and we thus again arrive at \eqref{constants_gen} (and thus $H=H(r)$).

However, in the special case $2p=n+2$ with $\B^2=0(\neq\E^2)$ (or $2p=n-2$ with $\E^2=0(\neq\B^2)$) one cannot in general conclude that $h(u,x)=U(u)X(x)$, and one has to consider the general equations \eqref{Ruu_2p=n+2}, \eqref{Ruu_2p=n-2}. If  $h(u,x)$ is indeed non-factorized, then this metric describes an Einstein space that admits a conformal (non-homothetic) map on Einstein spaces \cite{Ortaggio07,OrtPodZof08} and it is thus further constrained \cite{Brinkmann25} (in particular, it must be of constant curvature when $n=6$ {and $p=4$ or $p=2$ in the electric and magnetic case, respectively} \cite{OrtPodZof08}). One can still normalize $\R=0,\pm(n-2)(n-3)$.

\section{Summary for the generic case $2p\neq n$ ($n>4$): static black holes}

\label{sec_summary_gen}

\subsection{Case $2p\neq n\pm 2$: static black holes}

\label{subsec_summ_gen}

\subsubsection{Metric and Maxwell field}

\label{subsubsec_metric_generic}

Keeping in mind also the concluding observations of section~\ref{subsubsec_Ruu_gener}, the line-element is given by (cf. \eqref{geo_metric_text} with \eqref{f=0=e})
\be
  \d s^2=r^{2}h_{ij}\d x^i\d x^j{-}2\,\d u\d r-2H\d u^2 , \label{ds_generic}
\ee
where $h_{ij}=h_{ij}(x)$ is the metric of a Riemannian Einstein space of dimension $(n-2$) and scalar curvature $\R=K(n-2)(n-3)$, and (\eqref{grr_final} with \eqref{constants_gen})
\begin{eqnarray}
	& & 2H=K-\frac{2\Lambda}{(n-1)(n-2)}\,r^2-\frac{\mu}{r^{n-3}} \nonumber\\
	& & \qquad\qquad {}+\frac{\b}{n-2}\left[\frac{p-1}{n+1-2p}\frac{\E^2}{r^{2(n-p-1)}}-\frac{1}{p(n-1-2p)}\frac{\B^2}{r^{2(p-1)}}\right] \qquad (2p\neq n\pm1), \label{H_generic}
\end{eqnarray}
where $\Lambda$, $\mu$, $\E^2$, $\B^2$ and $K=0,\pm 1$ are all constants. The metric thus always admits the Killing vector field $\pa_u$, and it is static in regions where $H>0$, while roots of $H(r)$ define Killing horizons (see also \cite{BarCalCha12}). {Recall that when $2p=n\pm 1$ ($n$ odd), the second line of \eqref{H_generic} should be replaced by \eqref{H_log+} and \eqref{H_log-}, respectively.\footnote{However, it can be seen that for $n=5$, $p=3$ and $n=7$, $p=4$ necessarily $\E^2=0$, and for $n=5$, $p=2$ and $n=7$, $p=3$ necessarily $\B^2=0$ (see appendix~\ref{app_comments}) so that the logarithmic terms in \eqref{H_log+} and \eqref{H_log-} disappear in those cases. Therefore, effectively, logarithmic terms  in both \eqref{H_log+} and \eqref{H_log-}  can possibly arise only for an odd $n\ge9$.\label{footn_logs}}}

The ``Coulombian'' Maxwell field is given by (eqs. \eqref{F_aligned}, \eqref{Max0_r}, \eqref{Max0_ur} and \eqref{Fuigeneric})
\be
 \bF=\frac{1}{(p-2)!}\frac{1}{r^{n+2-2p}}\e_{i_1\ldots i_{p-2}}(x)\,\d u\wedge\d r\wedge\d x^{i_1}\wedge\ldots\wedge\d x^{i_{p-2}}+\frac{1}{p!}\bb_{i_1\ldots i_{p}}(x)\,\d x^{i_1}\wedge\ldots\wedge\d x^{i_{p}} , \label{F_generic}
\ee
where $\e_{i_1\ldots i_{p-2}}$ and $\bb_{i_1\ldots i_{p}}$ are harmonic forms (of respective rank $(p-2)$ and $p$) in the transverse geometry $h_{ij}$, i.e., they obey the Euclidean source-free Maxwell equations in $(n-2)$ dimensions (cf.~\eqref{diverg_elec} and the first and third of \eqref{diverg_magn}). These forms are, however, further constrained by the conditions \eqref{eqE}, \eqref{eqB} on the (constant) ``square'' of the field strengths, i.e.,
\be
 \E^2_{ij}=\frac{\E^2}{n-2}h_{ij} \quad {(p\ge 3)} , \qquad\qquad \B^2_{ij}=\frac{\B^2}{n-2}h_{ij} \quad {(p\le n-2)} . \label{EB_generic}
\ee
It is worth emphasizing again that conditions \eqref{EB_generic} do not only constrain the Maxwell field, but also impose severe restrictions on the Einstein metric $h_{ij}$. For instance, when $p=2$ it must also be almost-K\"ahler if $\B^2\neq0$ and $n$ must be even \cite{OrtPodZof08}. See \cite{BarCalCha12} and appendix \ref{app_comments} for further comments.

The above solutions can be seen as an extension to arbitrary $p$ of the $p=2$ ($n\neq6$) solutions studied in \cite{OrtPodZof08} (including, when $\B^2=0$, the higher-dimensional Reissner-Nordstr\"{o}m solution found by Tangherlini \cite{Tangherlini63}) and possess similar qualitative features. In particular, they represent static black holes (at least for certain values of the parameters in \eqref{H_generic}) dressed with electric and magnetic Maxwell fields. These solutions were previously obtained (starting from a static ansatz) and analyzed (including their thermodynamics) in \cite{BarCalCha12}, so that a detailed discussion is not necessary here. {Recalling that in an $(n-2)$-dimensional compact Riemannian space of positive constant curvature there exist no non-zero harmonic forms (except for a 0-form or a $(n-2)$-form) \cite{Bochner48,YanBocbook}, we observe that the metric $h_{ij}$ in \eqref{ds_generic} cannot describe a round sphere, except when $p=2$ and $\bb_{i_1 i_{2}}=0$ or, dually, when $p=n-2$ and $\e_{i_1\ldots i_{n-4}}=0$. Therefore, these static black holes cannot have a spherical horizon and cannot be asymptotically flat (in agreement with  \cite{Guven89,EmpOhaShi10,ShiOhaTan11}), except in the electric $p=2$ (or magnetic $p=n-2$) Reissner-Nordstr\"{o}m solution of \cite{Tangherlini63}. A flat and compact horizon metric $h_{ij}$ is instead permitted (giving $K=0$ in \eqref{H_generic}), in which case the harmonic forms $\e_{i_1\ldots i_{p-2}}$ and $\bb_{i_1\ldots i_{p}}$ must be covariantly constant \cite{Bochner48,YanBocbook}. This allows for, e.g., asymptotically locally (A)dS black holes (see also \cite{BarCalCha12}).}

Similarly as for the case $p=2$ \cite{OrtPodZof08}, an additional ``Vaidya-type'' matter field representing pure radiation aligned with $\bk$ (i.e., adding an extra term to the component $T_{uu}$ only) can easily be included by appropriately \cite{OrtPodZof08} modifying \eqref{Ruu_generic} and thus allowing for $\mu_{,u}\neq 0$ (see also a comment at the end of section~6 of \cite{BarCalCha12}) --- in the special case $2p=n$ this pure radiation can be sourced by the Maxwell field itself, as we show below in section~\ref{sec_2p=n}.

\subsubsection{{Weyl and Maxwell types}}

\label{subsubsec_algebr}

As discussed in \cite{OrtPodZof08}, the warped product structure of the metric \eqref{ds_generic} with $H=H(r)$ implies \cite{PraPraOrt07,HerOrtWyl13} (see also \cite{OrtPraPra13rev} for a compact summary of these results) that the corresponding Weyl type is D(bd) and that
\be
 \bk=\pa_r , \qquad \bl={-}\pa_u+H\pa_r
 \label{nulldirections}
\ee
are the (unique) double WANDs (no other WANDs --- not even single ones --- exist since the type is D(bd), {cf. appendix~\ref{app_Weyl}, in particular footnote~\ref{footn_WANDs};} $\bl$ also defines a second Robinson--Trautman null direction, as follows by ``time-reflection'' symmetry \cite{PraPraOrt07,OrtPodZof08}). Since the type is D(bd), all the Weyl components are uniquely determined from {(eqs.~\eqref{weyl0}--\eqref{weyl-2} of appendix~\eqref{app_summaryRT} with \eqref{H_generic})}
\beqn
 & & \Phi=-(n-2)(n-3)\frac{\mu}{2r^{n-1}}+\b\frac{n-3}{(n-1)(n-2)} \nonumber \\
 & & \qquad\qquad\qquad {}\times\left[\frac{(n-p)(2n-2p-1)(p-1)}{n+1-2p}\frac{\E^2}{r^{2(n-p)}}-\frac{2p-1}{n-1-2p}\frac{\B^2}{r^{2p}}\right] \qquad (2p\neq n\pm 1), \nonumber \label{Phi} \\
 & & \tilde\Phi_{\hat i\hat j\hat k\hat l}=r^{-2}{\cal C}_{\bar i\bar j\bar k\bar l} , \label{Weyl_generic}
\eeqn
where ${\cal C}_{\bar i\bar j\bar k\bar l}$ are the components of the Weyl tensor associated with $h_{ij}$ {in a frame of $h_{ij}$}, and the notation of \cite{Durkeeetal10,OrtPraPra13rev} is employed (with a hat denoting components {in a frame of the full spacetime metric $g_{\mu\nu}$, cf. appendix~\ref{app_Weyl}}). Since $h_{ij}$ must be Einstein, ${\cal C}_{\bar i\bar j\bar k\bar l}=0$ iff $h_{ij}$ is of constant curvature (which is necessarily the case when $n=5$), in which case the Weyl type becomes D(bcd). The scalar invariant $C_{\mu\nu\rho\sigma}C^{\mu\nu\rho\sigma}=4\Phi^2(n-1)/(n-3)+r^{-4}{\cal C}_{\bar i\bar j\bar k\bar l}{\cal C}_{\bar i\bar j\bar k\bar l}$ (cf. eqs.~(69) and (70) of \cite{OrtPraPra09}) also signals a curvature singularity at $r=0$.
{When $2p=n\pm 1$, the equation replacing \eqref{Phi} can be obtained by recalling \eqref{H_log+}, \eqref{H_log-} and using \eqref{weyl0}.}

$\bk$ and $\bl$ are manifestly also aligned null directions of the Maxwell field \eqref{F_generic} ($\bk$ is such by construction, recall \eqref{F_aligned}, and $\bl$ then due to (\ref{Fuigeneric})), which is thus also of type D. It follows that also the Ricci tensor is of (aligned) type D (as can be explicitly verified thanks to $R_{ui}=0$ and $R_{uu}= 2HR_{ur}$, cf. sections~\ref{subsec_Rui} and \ref{subsubsec_Ruu_gener}) --- apart from \eqref{nulldirections}, no other Ricci aligned null directions exist, not even single ones, as can be seen recalling that $R_{ij}\propto h_{ij}$. One can easily see that in a frame parallelly transported along $\bk$ (and adapted to \eqref{nulldirections}, {see appendix~\ref{app_Weyl}}) the electric and magnetic components of \eqref{F_generic} fall off, respectively, {with the monopole rate} $1/r^{n-p}$ and $1/r^{p}$ (as one could expect from a study of test fields \cite{Ortaggio14}).

Examples with $p=2$ (or, dually, $p=n-2$) are given, e.g., in \cite{Tangherlini63,OrtPodZof08}. Several other examples have been constructed in \cite{BarCalCha12}. Using a construction described in \cite{BarCalCha12} (see appendix~\ref{app_comments} for a brief summary) one can produce more solutions, as we now exemplify.

{\paragraph{Example  ($n=11$, $p=3 (8)$)} A magnetic solution with $n=11$, $p=3$ (or electric with $p=8$ after dualization) and a direct product transverse space is given by
\beqn
 & & h_{ij}=h^{(1)}_{ij}+h^{(2)}_{ij}+h^{(3)}_{ij} , \qquad 2H=K-\frac{\Lambda}{45}r^2-\frac{\mu}{r^8}-\frac{\b}{108}{\frac{{\B}^2}{r^4}} , \nonumber \\
 & & F_{123}=F_{456}=F_{789}=\frac{\B}{3\sqrt{2}} , \label{mag_n=11_p=3}
\eeqn
where $h^{(1)}_{ij}$, $h^{(2)}_{ij}$ and $h^{(3)}_{ij}$ are the metrics of three 3-dimensional spaces of constant curvature paramete\-rized, respectively, by the coordinates $(x_1,x_2,x_3)$, $(x_4,x_5,x_6)$ and $(x_7,x_8,x_9)$, and all of them having (constant) sectional curvature $4K$ (with $K=0,\pm1$). {For a suitable choice of the parameters this represents magnetic static black holes with a direct product horizon.} (Of course this solution can be dualized to an electric form with $p=8$.)}

\subsection{Exceptional case $2p=n\pm 2$ ($n\ge6$, even)}

\label{subsubsec_summary_2p=n+2}

These special ranks of $\bF$ also fall into the discussion of section~\ref{subsec_summ_gen} if the \emph{additional} assumptions $(\E^2)_{,i}=0$ (for $2p=n+2$) or $(\B^2)_{,i}=0$ (for $2p=n-2$) are made (or if the transverse space is assumed to be compact and the electric and magnetic fields are both non-zero, {cf. section~\ref{subsubsec_Ruu_2p=n+2}}), in which case they again describe static black holes, as in the $n=6$, $p=2$  (or $p=4$ after dualization) example given by eq.~(66) of \cite{OrtPodZof08} (with $D=6$), or as in the following example.

\paragraph{Example  ($n=8$, $p=5 (3)$)} A solution describing an electric field with $n=8$, $p=5$ (or a magnetic field with $p=3$ after dualization) is
\beqn
 & & h_{ij}=\delta_{ij} , \qquad 2H=-\frac{\Lambda}{21}r^2-\frac{\mu}{r^5}-\frac{\b}{3r^4}2{\E}^2 , \nonumber \\
 & & F_{ur123}=F_{ur456}=\frac{\E}{2\sqrt{3}} . \label{el_n=8_p=5}
\eeqn
Note that here the electric field is $r$-independent since $2p=2+n$ (first term of \eqref{F_generic}). {These are locally AdS (for $\Lambda<0$) electric static black holes with a flat horizon.}

 However, more general solutions are now permitted for $2p=n+2$ with $\B^2=0$, {$\E^2=\E^2(u,x)$} or for $2p=n-2$ with $\E^2=0$, {$\B^2=\B^2(u,x)$} (or for a non-compact transverse space).
The line-element is again given by \eqref{ds_generic}, but $H$, given by \eqref{grr_final}, can generically depend on all coordinates {(and the metric is thus {\em non-static}, in general)}. The Einstein metric $h_{ij}=h_{ij}(u,x)$ is {generically non-factorized, and thus} further constrained by the property of admitting conformal maps on Einstein spaces \cite{Brinkmann25}. The Maxwell field is still given by \eqref{F_generic} and satisfies the source-free Maxwell equations in the transverse geometry $h_{ij}$, but {$\E^2$ (or $\B^2$) can now depend on $x$ and $u$ (eqs.~\eqref{EB_u}); the form $\e_{i_1\ldots i_{p-2}}$ (or $\bb_{i_1\ldots i_{p}}$) is still $u$-independent (eq. \eqref{e_u} with $n=2p+2$, and \eqref{b_u}).} The remaining equations to be satisfied are \eqref{Ruu_2p=n+2}, \eqref{Ruu_2p=n-2} (see section~\ref{subsubsec_Ruu_2p=n+2} for more comments). Because \eqref{F_generic} holds also here, the {Maxwell and} Ricci tensors obviously possess the same algebraic properties as discussed in section~\ref{subsec_summ_gen} and are thus still of type D, doubly aligned with both null vectors \eqref{nulldirections}. The Weyl type is {generically II(bd) (see section~\ref{subsec_Rur} and appendix~\eqref{app_Weyl})} and the Weyl components of b.w. 0 (where $\Phi$ is necessarily non-zero) are still given by \eqref{Weyl_generic}. {We observe that $H_{,i}\neq0$ implies that} the vector $\bl$ is non-geodesic (appendix~\ref{app_Weyl}) {and not a multiple WAND, as can be seen using the results of appendix~\ref{app_l_mWAND} with \eqref{grr_final} and \eqref{Ruu_2p=n+2}, \eqref{Ruu_2p=n-2} (and vice versa, i.e., if $H_{,i}=0$ then $\bl$ is a geodesic multiple WAND, as observed in section~\ref{subsubsec_algebr} for the case $2p\neq n\pm2$).}

It should also be observed that the ranks $2p=n\pm 2$ are special (even in the subcase of static metrics) because {contributions from the electric and magnetic terms to the energy-momentum components $T_{ij}$ cancel out (thanks to \eqref{eqE}, \eqref{eqB}) for $2p=n+2$ and for $2p=n-2$, respectively. Related to this, for even $p$ there can exist forms $\e_{i_1\ldots i_{p-2}}$ (for $2p=n+2$) or $\bb_{i_1\ldots i_{p}}$ (for $2p=n-2$) that are self-dual in the transverse geometry $h_{ij}$, cf. also appendix~\ref{app_comments}.}

\section{Special case $2p=n$ ($n$ even): black holes with electromagnetic radiation}

\label{sec_2p=n}

{As we observed, the rank satisfying $2p=n$ has special properties and needs to be studied separately. This is not so surprising since this is the unique rank for which the Maxwell equations are conformally invariant and admit self-dual solutions (for odd $p$) or a continuous duality symmetry (for even $p$), cf., e.g., \cite{HenTei86,HenTei88,Deseretal97,Bremeretal98}.}

The results of sections \ref{subsec_Max1} and \ref{subsec_Rur} are still valid also in the case $2p=n$ (but recall \eqref{eqEB}). Instead, in sections \ref{subsec_Max2}, \ref{subsec_Rui}  and \ref{subsubsec_Ruu_gener} we assumed $2p\neq n$ and those results are modified as follows.

\subsection{Remaining field equations for the case $2p=n$}

\label{subsec_integr_2p=n}

\subsubsection{Maxwell equations, step two}

\label{subsubsec_Max2_spec}

Certain terms in the Maxwell equations studied in section~\ref{subsec_Max2} combine since they have the same $r$-dependence, and the equations of section~\ref{subsec_Max2} are thus replaced by\footnote{For later purposes it may be useful to observe that in an index-free notation the first of these equations reads $\star\,\d\!\star\!{\mbox{\boldmath{$b$}}}=-\d{\mbox{\boldmath{$e$}}}$, where $\star$ is the Hodge dual in the transverse space of $h_{ij}$ (not to be confused with the $n$-dimensional Hodge dual $^*$ in the full spacetime geometry $g_{\mu\nu}$, defined in section~\ref{intro}).\label{footn_star}}
\beqn
    & & {(\sqrt{h}\, \bb^{kj_1 \ldots j_{p-1}})_{,k}}= \frac{1}{2}(n-2)\sqrt{h}\, h^{i_1 j_1} \ldots h^{i_{p-1} j_{p-1}} \e_{[i_2 \ldots i_{p-1}, i_1]} , \label{maxw_2p_1} \\
		& & \bb_{i_1 \ldots i_{p},u}=\frac{n}{2}{f}_{[i_2 \ldots i_{p},i_1]} , \label{b_u_2p} \\
		& & {(\sqrt{h}\, \e^{i_1 \ldots i_{p-2}})_{,u} = (\sqrt{h}\, {f}^{k i_1 \ldots i_{p-2}})_{,k} .} \label{maxw_2p_2}
\eeqn
One also obtains an additional equation $(\sqrt{h}\, h^{kl} h^{i_1 j_1} \ldots h^{i_{p-2} j_{p-2}} \e_{[i_1 \ldots i_{p-2},k]})_{,l} = 0$ {($\Leftrightarrow \star\,\d\!\star\!\d{\mbox{\boldmath{$e$}}}=0$)}, which is however identically satisfied by virtue of \eqref{maxw_2p_1} and the antisymmetry of $\bb_{i_1 \ldots i_{p}}$.

Note that the r.h.s. of \eqref{maxw_2p_1} acts as a ``current'' for the $p$-form $\bb_{i_1 \ldots i_{p}}$ so that $\bb_{i_1 \ldots i_{p}}$ and $\e_{[i_2 \ldots i_{p-1}, i_1]}$ no longer satisfy the Euclidean source-free Maxwell equations in the transverse space. A fundamental difference from the generic case $2p\neq n$ is that now the components $F_{u i_1 \ldots i_{p-1}}$ (eq.~\eqref{Max-1_r}) can be non-zero. Instead of~\eqref{EB_u} we now obtain from \eqref{maxw_2p_2} and \eqref{b_u_2p} (with \eqref{h_u})
\beqn
 & & (n-2)(\E^2)_{,u}=-n\E^2(\ln\sqrt{h})_{,u}{+}{(n-2)\frac{2}{\sqrt{h}}\,e_{i_1\ldots i_{p-2}}(\sqrt{h}\, {f}^{k i_1 \ldots i_{p-2}})_{,k}}, \label{E_u_2p=n} \\
 & & (n-2)(\B^2)_{,u}=-n\B^2(\ln\sqrt{h})_{,u}+n(n-2)b^{i_1i_2\ldots i_p}{f}_{[i_2 \ldots i_{p},i_1]} . \label{B_u_2p=n}
\eeqn

\subsubsection{Einstein equation for $R_{ui}$}

\label{Rui_2p=n}

Since in general $F_{u i_1 \ldots i_{p-1}}\neq0$, also ${F_{u\a_1\ldots\a_{p-1}} {F_i}^{\a_1\ldots\a_{p-1}}\not=0}$ and the Einstein equation~\eqref{Ricci} for $R_{ui}$ thus becomes
\be
   R_{ui} =\frac{\b}{4}\left[{-}(n-2)\,{\delta}_{i}^{\ k}\e^{j_1\ldots j_{p-2}}-2\bb_i^{\ kj_1\ldots j_{p-2}}\right]\Big[r^{1-n}(n-2)\e_{[j_1\ldots j_{p-2},k]}-2r^{2-n}{f}_{kj_1\ldots j_{p-2}}\Big] .
\ee
Instead of \eqref{constrijsi}--\eqref{eqBi} this now gives
\beqn
  & & \mu_{,i} = \b\left[(n-2)\,{\delta}_{i}^{\ k}\e^{j_1\ldots j_{p-2}}{+}2\bb_i^{\ kj_1\ldots j_{p-2}}\right]{f}_{kj_1\ldots j_{p-2}} , \label{mu_i_2p=n} \\
  & & \left(\E^2+\frac{4}{n(n-2)}\B^2\right)_{,i}=\frac{n-2}{2}\left[(n-2){\delta}_{i}^{\ k}\e^{j_1\ldots j_{p-2}}{+}2\bb_i^{\ kj_1\ldots j_{p-2}}\right]\e_{[j_1\ldots j_{p-2},k]} ,  \label{EB_i_2p=n}
\eeqn
while $(n-4)\R_{,i}=0$ still holds.

\subsubsection{Einstein equation for $R_{uu}$}

\label{Ruu_2p=n}

When $2p=n$ this equation is different for various reasons. First, since generically $F_{u i_1 \ldots i_{p-1}}\neq0$, the r.h.s of \eqref{Ruu_eq} is not zero but
\beqn
 & & \b h^{i_1j_1}\ldots h^{i_{p-1}j_{p-1}}\bigg[r^{-n}\frac{(n-2)^2}{4}\e_{[i_2\ldots i_{p-1},i_1]}\e_{[j_2\ldots j_{p-1},j_1]}  \nonumber \\
 & & {\qquad}+r^{2-n}{f}_{i_1\ldots i_{p-1}}{f}_{j_1\ldots j_{p-1}}-r^{1-n}(n-2)\e_{[i_2\ldots i_{p-1},i_1]}{f}_{j_1\ldots j_{p-1}}\bigg] .  \label{Ruu_eq_2p=n}
\eeqn
Further, we now generically have $\triangle\E\neq 0$, $\triangle\B\neq 0$, $\triangle\mu\neq 0$ (cf. section~\ref{Rui_2p=n}), and \eqref{EB_u} are not satisfied (section~\ref{subsubsec_Max2_spec}), giving rise to further terms of order $r^{-n}$ and $r^{1-n}$ in $R_{uu}$ (cf. the l.h.s. of \eqref{Ruu_eq}). Keeping \eqref{Ruu_eq_2p=n} and all these terms in mind, instead of \eqref{Ruu_generic} one obtains (at order $r^{2-n}$, $r^{-n}$, $r^{1-n}$, respectively)
\beqn
 & & (n-2)\mu_{,u}=-(n-1)\mu(\ln\sqrt{h})_{,u}{-}2\b\F^2 \qquad (n>4) ,  \label{mu_u_2p=n} \\
 & & \triangle\mu=\frac{\b}{2}\left[(n-2)\left(\E^2+\frac{4}{n(n-2)}\B^2\right)_{,u}+n\left(\E^2+\frac{4}{n(n-2)}\B^2\right)(\ln\sqrt{h})_{,u}\right] \nonumber \\
 & &  \qquad\qquad {}+2\b(n-2)\,\e_{[i_2\ldots i_{p-1},i_1]}\,{f}^{i_1\ldots i_{p-1}} , \label{Laplac_mu_2p=n} \\
 & & \triangle \left(\E^2+\frac{4}{n(n-2)}\B^2\right)=(n-2)^2\, h^{i_1j_1}\ldots h^{i_{p-1}j_{p-1}}\e_{[i_2\ldots i_{p-1},i_1]}\,\e_{[j_2\ldots j_{p-1},j_1]} , \label{Laplac_EB_2p=n}
\eeqn
where we have defined the $r$-independent quantity
\be
 \F^2\equiv h^{i_1j_1}\ldots h^{i_{p-1}j_{p-1}}{f}_{i_1\ldots i_{p-1}}{f}_{j_1\ldots j_{p-1}}{={f}^{i_1\ldots i_{p-1}}{f}_{i_1\ldots i_{p-1}}} .
\label{F2}
\ee
{We observe that {\em \eqref{Laplac_mu_2p=n} and \eqref{Laplac_EB_2p=n}  are identically satisfied} by virtue of \eqref{mu_i_2p=n} and \eqref{EB_i_2p=n}, respectively (after using \eqref{maxw_2p_1}--\eqref{B_u_2p=n}) --- this will be understood from now on.}
Note that if $\F^2=0$ ($\Leftrightarrow {f}_{i_1\ldots i_{p-1}}=0$) then \eqref{mu_i_2p=n} gives $\mu_{,i} = 0$ and thus $\mu(\ln\sqrt{h})_{,ui}=0$ by \eqref{mu_u_2p=n}, which implies (as explained in section~\eqref{subsubsec_Ruu_gener}) that if $\mu\neq 0$ one can rescale the coordinates so as to have $h_{ij}=h_{ij}(x)$, $\R=\mbox{const}$, $\mu=\mbox{const}$. Vice versa, $\F^2\neq0$ clearly requires that $\mu_{,u}$ and $(\ln\sqrt{h})_{,u}$ cannot both vanish simultaneously, so that $\pa_u$ cannot be a Killing vector field (as opposed to the case of static black holes, cf. section~\ref{subsubsec_metric_generic}). For $n=4$ ($p=2$) the l.h.s. of \eqref{mu_u_2p=n} contains an additional term ${-}\frac{1}{2}\triangle{\cal R}$ \cite{RobTra62,Stephanibook} (cf. also appendix~B of \cite{OrtPodZof08}\footnote{Eq.~(B.13) of \cite{OrtPodZof08} contains a typo: its r.h.s. should read $8P^2(Q_{,1}\xi_1 + Q_{,2}\xi_2)$.}) and this argument does not apply.

\subsection{Summary and discussion}

\label{subsec_2p=n_summary}

We first observe that static black hole solutions clearly exist also for the special rank $2p=n$ (see section~\ref{subsec_D_Max} below), to which a discussion similar to that of section~\ref{subsec_summ_gen} still applies. However, there exists now also a new family of solutions in the case $F_{u j_1 \ldots j_{p-1}}\neq 0$. The Maxwell field is given by
\beqn
 \bF= & & \frac{1}{\left(\frac{n}{2}-2\right)!}\frac{1}{r^{2}}\e_{i_1\ldots i_{p-2}}\d u\wedge\d r\wedge\d x^{i_1}\wedge\ldots\wedge\d x^{i_{p-2}}+\frac{1}{\left(\frac{n}{2}\right)!}\bb_{i_1\ldots i_{p}}\d x^{i_1}\wedge\ldots\wedge\d x^{i_{p}} \nonumber \\
			& & {}+\frac{1}{2\left(\frac{n}{2}-1\right)!}\left(-\frac{n-2}{r}\e_{[i_2\ldots i_{p-1},i_1]}+2{f}_{i_1\ldots i_{p-1}}\right)\d u\wedge\d x^{i_1}\wedge\ldots\wedge\d x^{i_{p-1}} . \label{F_2p=n}
\eeqn
The forms $\e_{i_1\ldots i_{p-2}}(u,x)$ and $\bb_{i_1\ldots i_{p}}(u,x)$ are generally not harmonic in the transverse space, but satisfy the ``modified'' Euclidean Maxwell equations in $(n-2)$ dimensions \eqref{diverg_elec} and \eqref{maxw_2p_1}. In addition, they can depend on $u$, cf. eqs. \eqref{b_u_2p}, \eqref{maxw_2p_2}. The latter further tells us that the $(p-1)$-form ${f}_{i_1\ldots i_{p-1}}(u,x)$ is also generically non-harmonic. Notice that the Maxwell equations do not specify the $u$-dependence of ${f}_{i_1\ldots i_{p-1}}$, which can be interpreted as a freedom in the choice of a ``wave profile''.

The line-element is given by \eqref{ds_generic} with
\be 2H=K{+}\frac{2(\ln\sqrt{h})_{,u}}{n-2}\,r-\frac{2\Lambda}{(n-1)(n-2)}\,r^2-\frac{\mu}{r^{n-3}} +\frac{\b}{2}\left(\E^2+\frac{4}{n(n-2)}\B^2\right)\frac{1}{r^{n-2}} , \label{grr_2p=n}
\ee
where $\E^2$, $\B^2$ and $\mu$ are generically functions of $u$ and $x$, cf. the corresponding equations  \eqref{E_u_2p=n}, \eqref{B_u_2p=n}, \eqref{mu_i_2p=n}, \eqref{EB_i_2p=n}, \eqref{mu_u_2p=n}. The metric $h_{ij}(u,x)=h^{1/(n-2)}(u,x)\,\gamma_{ij}(x)$ is again Einstein, with scalar curvature $\R=K(n-2)(n-3)$ (where $K=0,\pm 1$). One further has the constraint \eqref{eqEB}.

If $h_{ij}$ is taken to be the metric of a \emph{compact} {space}, then by \eqref{Laplac_EB_2p=n} necessarily \cite{Bochner48,YanBocbook,KobNom2} $\e_{[i_2\ldots i_{p-1},i_1]}=0$ (since the r.h.s of \eqref{Laplac_EB_2p=n} is non-negative), so that $\left(\E^2+\frac{4}{n(n-2)}\B^2\right)_{,i}=0$ (cf.~\eqref{EB_i_2p=n}), and $\e_{i_1\ldots i_{p-2}}$ and $\bb_{i_1\ldots i_{p}}$ must both be harmonic (eqs.~\eqref{diverg_elec}, \eqref{maxw_2p_1}). In particular, it follows \cite{Bochner48,YanBocbook} that {\em $h_{ij}$ cannot describe a round sphere}, and no asymptotically flat spacetimes are thus to be found in this class of solutions (note that if $\e_{i_1\ldots i_{p-2}}=0=\bb_{i_1\ldots i_{p}}$, as in section~\ref{subsec_null_Max} below, the argument still applies since in that case it is ${f}_{i_1\ldots i_{p-1}}$ that must be harmonic, cf.~\eqref{b_u_2p}, \eqref{maxw_2p_2}).

These solutions also include the standard electrovac Robinson--Trautman solutions with $2p=n=4$, provided an extra term proportional to $\triangle{\cal R}$ is incorporated into \eqref{mu_u_2p=n} \cite{RobTra62,Stephanibook,OrtPodZof08}.

\subsubsection{Maxwell type and self-duality}

The Maxwell field \eqref{F_2p=n} is in general of type II (aligned with $\bk$ by construction) and, in a parallelly transported frame adapted to \eqref{nulldirections} {(appendix~\ref{app_Weyl})}, it peels off as (in agreement with test-field results \cite{Ortaggio14})
\be \bF=\frac{\mbox{\boldmath{$N$}}}{r^{\frac{n}{2}-1}}+\frac{\mbox{\boldmath{$II$}}}{r^{\frac{n}{2}}} ,
	\label{peeling_n/2}
\ee
{where the symbols $\mbox{\boldmath{$II$}}$ and $\mbox{\boldmath{$N$}}$ specify the algebraic type of the corresponding terms,} with the ``radiative'' $\mbox{\boldmath{$N$}}$ term proportional to ${f}_{i_1\ldots i_{p-1}}$. {The Maxwell type becomes D when there exists a second null direction aligned with $\bF$, i.e. (using a null rotation \eqref{nullrot}), iff the following equation admits a solution\footnote{After contraction with $z_{\bar i_{1}}$ eq.~\eqref{Maxw_D} gives for $n>4$ (i.e., $p>2$) the necessary condition ${f}_{\bar i_1\ldots \bar i_{p-1}}z_{\bar i_{1}}=0$, which generically will not be satisfied, thus showing that  the generic Maxwell type is indeed II here (while for $n=4$ ($p=2$) the Maxwell types II and D are always equivalent, cf., e.g., \cite{Stephanibook}).}
for the null rotation parameter $z_{\bar i_{p}}\equiv r^{-1}z_{\hat i_{p}}$
\be
 -\frac{n-2}{2}\e_{[\bar i_2\ldots \bar i_{p-1},\bar i_1]}+r\,{f}_{\bar i_1\ldots \bar i_{p-1}}=-\frac{n-2}{2}\,r\,\e_{[\bar i_2\ldots \bar i_{p-1}}z_{\bar i_1]}{+}r\,z_{\bar i_{p}}\bb_{\bar i_{p}\bar i_1\ldots \bar i_{p-1}} ,
\label{Maxw_D}
\ee
see solution~\eqref{metric_n=8_p=4_D}  below for an example. As a special case, this occurs (with $z_{\bar i_{1}}=0$) when} the frame vector $\bl$ \eqref{nulldirections} is aligned with the Maxwell field, i.e., iff ${f}_{i_1\ldots i_{p-1}}=0=\e_{[i_2\ldots i_{p-1},i_1]}$ (which implies that the radiative term vanishes).  On the other hand, the Maxwell type is N iff {$\bk$ is doubly aligned, i.e.,} $\e_{i_1\ldots i_{p-2}}=0=\bb_{i_1\ldots i_{p}}$ --- these special cases are discussed below in sections~\ref{subsec_D_Max} and \ref{subsec_null_Max}. We observe that for $n=6$ ($p=3$) only the Maxwell type N is possible (cf. appendix~\ref{app_comm_special}). {Let us further notice that the b.w. -2 component $8\pi T_{\mu\nu} l^\mu l^\nu=R_{\mu\nu} l^\mu l^\nu{=R_{uu}{-}2HR_{ur}}$ {(with \eqref{nulldirections} and $R_{rr}=0$)} of the energy-momentum tensor, {representing the flux of electromagnetic energy along $\bk$,} equals expression~\eqref{Ruu_eq_2p=n} {and is characterized by the} leading term $\b r^{2-n}\F^2$ {--  an invariant quantity taking} the same value in any frame {parallelly transported along} $\bk$ (i.e., it is invariant under a null rotation \eqref{nullrot} of the frame \eqref{frame} with $z_{\hat i}$ independent of $r$, {and under spins}).}

When $p$ is {\em odd}, straightforward calculations show that the field \eqref{F_2p=n} is {\em self-dual} ($^*\bF=\bF$) iff ${\mbox{\boldmath{$e$}}}=\star{\mbox{\boldmath{$b$}}}$, $\star\d{\mbox{\boldmath{$e$}}}=-\d{\mbox{\boldmath{$e$}}}$ and $\star{\mbox{\boldmath{$f$}}}=-{\mbox{\boldmath{$f$}}}$ (recall the definition of $\star$ in footnote~\ref{footn_star}), whereas self-duality is not possible for an even $p$ \cite{HenTei88} (in particular, the condition ${\mbox{\boldmath{$e$}}}=\star{\mbox{\boldmath{$b$}}}$ implies $p(p-1)\E^2=\B^2$, and that the l.h.s. and r.h.s. of \eqref{eqEB} must vanish separately).  Examples satisfying these conditions (with $\E^2=0=\B^2$) are given by \eqref{6D_null} and \eqref{6D_null_K} (under the conditions described there).

\subsubsection{Weyl type}

In general, the Weyl tensor is of type II(bd), see section~\ref{subsec_Rur} and appendix~\ref{app_Weyl}. The nontrivial Weyl components read (cf. \eqref{weyl0}--\eqref{weyl-2} with \eqref{grr_2p=n})
\beqn
 & & \Phi=-\frac{1}{2}(n-2)(n-3)\,\mu\,r^{1-n}+\frac{1}{4}n(n-3)\,\b\!
 \left(\E^2+\frac{4}{n(n-2)}\, \B^2\right)r^{-n} ,  \label{weyl0_text} \\
 & & \tilde\Phi_{\hat i\hat j\hat k\hat l}= r^{-2}{\cal C}_{\bar i\bar j\bar k\bar l} , \label{ijkl_text} \\
 & & \Psi'_{\hat i}=m_{(\hat i)}^k\frac{n-3}{n-2}\left[
 \frac{1}{2}(n-1)\,\mu_{,k}\,r^{1-n}-\frac{1}{4}n\,\b\!
 \left(\E^2+\frac{4}{n(n-2)}\, \B^2\right)_{,k}\,r^{-n}
 \right] , \label{weyl-1_text} \\
 & & \Omega'_{\hat i\hat j}=m_{(\hat i)}^k m_{(\hat j)}^l
 \Bigg\{
 -\frac{1}{2}\!\left(\mu_{||kl}-\triangle\mu\,\frac{h_{kl}}{n-2}\right)r^{1-n}  \nonumber\\
 & &\hspace{23mm}
 +\frac{1}{4}\b\!\left[\left(\E^2+\frac{4}{n(n-2)}\, \B^2\right)_{||kl}-
 \triangle \left(\E^2+\frac{4}{n(n-2)}\, \B^2\right)\!\frac{h_{kl}}{n-2}\right]r^{-n}
 \Bigg\} , \label{weyl-2_text}
\eeqn
where {\em \eqref{weyl-2_text} holds only for $n>4$} since the identity \eqref{identity_||} has been employed (see \cite{Stephanibook} for $n=4$), {and the terms appearing in (\ref{weyl-1_text}) and (\ref{weyl-2_text}) are determined by (\ref{mu_i_2p=n}), (\ref{EB_i_2p=n}) (and (\ref{Laplac_mu_2p=n}), (\ref{Laplac_EB_2p=n})).}
Eqs.~\eqref{weyl0_text} and \eqref{ijkl_text} (which coincide with \eqref{Weyl_generic} for $2p=n$) imply that $\bk$ can never be a triple (or quadruple) WAND, since this would require $\mu=\E^2=\B^2={\cal C}_{\bar i\bar j\bar k\bar l}=0$ (and thus ${f}_{i_1\ldots i_{p-1}}=0$ thanks to \eqref{mu_u_2p=n}), leading to a vacuum spacetime of constant curvature --- except for this trivial case, the Weyl types III, N and O are thus \emph{forbidden}. The Weyl type D(bd) (or D(bcd)) is possible in special cases, for example for metrics \eqref{metric_n=8_p=4_II} and \eqref{metric_n=8_p=4_II_2} below, when $\bl$ is aligned with $\bF$ (section~\ref{subsec_D_Max}), or when $\bk$ is doubly aligned with $\bF$ (section~\ref{subsec_null_Max}). As in section~\ref{subsubsec_algebr}, $C_{\mu\nu\rho\sigma}C^{\mu\nu\rho\sigma}=4\Phi^2(n-1)/(n-3)+r^{-4}{\cal C}_{\bar i\bar j\bar k\bar l}{\cal C}_{\bar i\bar j\bar k\bar l}$.

\subsubsection{Examples}

Some explicit examples of various Weyl and Maxwell types are provided below.

\paragraph{Example ($n=8$, $p=4$)}

It is easy to construct examples if the transverse space is taken to be flat, the forms $\e_{i_1\ldots i_{p-2}}$, $\bb_{i_1\ldots i_{p}}$ and ${f}_{i_1\ldots i_{p-1}}$ to be harmonic --- in fact $x$-independent --- and $\mu=\mu(u)$ {(with $(\E^2)_{,u}=0=(\B^2)_{,u}$ by \eqref{E_u_2p=n}, \eqref{B_u_2p=n}).} A simple example with $n=8$, $p=4$ is given by \eqref{ds_generic} and \eqref{F_2p=n} with (recall~\eqref{F2})
\beqn
 & & h_{ij}=\delta_{ij} , \quad 2H=-\frac{\Lambda}{21}r^2-\frac{\mu(u)}{r^5}+\frac{\b}{2r^6}\left({\E}^2+\frac{{\B}^2}{12}\right) , \quad \mu(u)=\mu_0{-}\frac{\b}{3}\int\F^2\d u , \label{metric_n=8_p=4_II} \nonumber \\
 & & F_{ur12}=F_{ur34}=F_{ur56}=\frac{\E}{\sqrt{6}}\frac{1}{r^2} , \qquad F_{1234}=F_{1256}=F_{3456}=\frac{\B}{6\sqrt{2}} , \\
 & & F_{uijk}=f_{ijk}(u) \qquad \mbox{ with } F_{ui12}=F_{ui34}=F_{ui56}=0 , \nonumber
\eeqn
where $\E$ and $\B$ are constants. This spacetime generically represents a collapse to (or evaporation of) black holes with a flat horizon in the presence of {an electromagnetic field which consists of both a static component and of $u$-dependent contracting (or expanding) radiation.}\footnote{It should be observed that no direct four-dimensional analog of such solutions exists since in 4D the presence of non-vanishing b.w.~0 components of the Maxwell field would imply (by \eqref{mu_i_2p=n}) that $\mu_{,i}\neq 0$ (cf.~(28.37e) of \cite{Stephanibook}, or (B.11) of \cite{OrtPodZof08}).} {Correspondingly, the mass parameter $\mu$ is $u$-dependent and monotonically increases (or decreases) according to the time-orientation of $\bk$ (thus corresponding to received or emitted radiation, cf.~\cite{Senovilla14}).} These metrics are asymptotically locally (A)dS if $\Lambda\neq 0$. When $\Lambda=0$, in regions where $\mu$=const$>0$ they clearly possess a horizon, but they are not asymptotically flat.\footnote{The line-element corresponding to~\eqref{metric_n=8_p=4_II} is static when $\mu$=const$>0$ and $H>0$, in which case the zeros of $H$ define Killing horizons. For the dynamical metrics with $\mu=\mu(u)$, instead of giving a detailed analysis of the parameters range which ensures the existence of marginally trapped tubes and dynamical horizons,  we refer to, e.g., \cite{Senovilla14,DadFirMan12}, where similar four-dimensional spacetimes have been discussed in more detail.} {The Weyl type is D(bcd) because here $H=H(u,r)$ and $h_{ij}$ is flat \cite{PraPraOrt07,HerOrtWyl13,OrtPraPra13rev} (as can also be seen explicitly from \eqref{weyl0_text}--\eqref{weyl-2_text}).} Generically, the Maxwell field is of genuine type II.  Here ${\E}$ and ${\B}$ are independent parameters and, in particular, can vanish independently (in other words, \eqref{eqE} and \eqref{eqB} are satisfied separately). However, since here $2p=n$, solutions exist that satisfy only the weaker constraint \eqref{eqEB} (i.e., the electric and magnetic fields must both be non-zero), as illustrated by the following example.

\paragraph{Example ($n=8$, $p=4$)}

To obtain a different solution with $n=8$, $p=4$, one can specify the metric functions and electromagnetic field by
\beqn
 & & h_{ij}=\delta_{ij} , \quad 2H=-\frac{\Lambda}{21}r^2-\frac{\mu(u)}{r^5}+\b\frac{3{\E}^2}{4r^6} , \quad \mu(u)=\mu_0{-}\frac{\b}{3}\int\F^2\d u , \label{metric_n=8_p=4_II_2} \nonumber \\
 & & F_{ur12}=F_{ur34}=\frac{\E}{2}\frac{1}{r^2} , \qquad F_{1234}=\frac{\E}{2} ,  \\
 & & F_{uijk}=f_{ijk}(u) \qquad \mbox{ with } F_{ui12}=0=F_{ui34} , \nonumber
\eeqn
where $\E$ is a constant. Clearly \eqref{eqE} and \eqref{eqB} are not satisfied in this case. If  $\E\neq 0\neq f_{ijk}$ the Maxwell field is of genuine type II. Comments similar to those given for \eqref{metric_n=8_p=4_II} apply also here, {in particular the Weyl type is again D(bcd).}

{\paragraph{Example ($n=8$, $p=4$)} An example in which $\mu$ has also a non-trivial $x$-dependence is given by
\beqn
 & & h_{ij}=\delta_{ij} , \quad 2H=-\frac{\Lambda}{21}r^2-\frac{\mu(u,x_1)}{r^5}+\frac{\b{\B}^2}{24r^6} , \quad \mu(u,x_1)=\mu_0-2\b f_0(\sqrt{2}\B x_1{+}2f_0u) , \label{metric_n=8_p=4_D} \nonumber \\
 & & F_{1234}=F_{1256}=F_{3456}=\frac{\B}{6\sqrt{2}} , \qquad F_{u234}=F_{u256}=f_0,
\eeqn
where $\B$ and $f_0$ are constants (for $f_0=0$ this trivially reduces to a subcase of \eqref{metric_n=8_p=4_II}). Here $\Phi\neq0\neq\Psi'_{\hat i}$ while $\tilde\Phi_{\hat i\hat j\hat k\hat l}=0=\Omega'_{\hat i\hat j}$ (see \eqref{weyl0_text}--\eqref{weyl-2_text}), so that the Weyl tensor is of genuine type II(bcd) (by the argument in appendix~\ref{app_type_D(bd)}, since here \eqref{z_cond_3_gen} admits no solutions). The vector field $\bl$ is clearly not aligned with $\bF$, nevertheless the Maxwell type is D, a second aligned null direction being given by a null rotation \eqref{nullrot} with $\bbm_{(\hat 1)}=r^{-1}\pa_{x_1}$ and the only non-zero parameter $z_{\hat 1}=-6\sqrt{2}K_0\B^{-1}r$ (cf.~\eqref{Maxw_D}).
}

\subsection{{$\bl$ aligned with $\bF$: static black holes ($n\ge 8$)}}

\label{subsec_D_Max}

{The null vector field $\bl$ of \eqref{nulldirections} is uniquely defined by certain geometrical properties (appendix~\ref{app_Weyl}). Furthermore, as noticed above, it is aligned with $\bF$ iff
\be
  {f}_{i_1\ldots i_{p-1}}=0 , \qquad \e_{[i_2\ldots i_{p-1},i_1]}=0 ,
\ee	
which implies that the Maxwell field is of type D and non-radiative (as mentioned above and in appendix~\ref{app_comm_special}, for $n=6$ this would lead to $\bF=0$, so we can restrict here to $n\ge8$) { --- indeed here $T_{\mu\nu} l^\mu l^\nu=0$.} Equations \eqref{maxw_2p_1}, \eqref{b_u_2p}, \eqref{maxw_2p_2}, \eqref{E_u_2p=n}, \eqref{B_u_2p=n}, \eqref{mu_i_2p=n} and \eqref{mu_u_2p=n} now take exactly the same form as in the generic case $2p\neq n$ (in particular, \eqref{E_u_2p=n}, \eqref{B_u_2p=n} reduce to \eqref{EB_u}). Instead of the conditions $\E^2_{,i}=0=\B^2_{,i}$ found in the $2p\neq n$ case, one has the weaker condition (from \eqref{EB_i_2p=n})
\be
	\left(\E^2+\frac{4}{n(n-2)}\B^2\right)_{,i}=0 .
	\label{EB_i_2p_D}
\ee
However, together with \eqref{EB_u} this suffices to show that, again, one can rescale away the $u$-dependence of $h_{ij}$ (and thus also of $\mu$ and $\e_{i_1\ldots i_{p-2}}$), and the discussion of section~\ref{subsec_summ_gen} then applies (with the only difference that \eqref{EB_generic} are replaced by \eqref{eqEB}, and that \eqref{EB_i_2p_D} replaces $\E^2_{,i}=0=\B^2_{,i}$). The metric is thus \eqref{ds_generic} with $H(r)$ given by \eqref{grr_2p=n} but without the $\frac{2}{n-2}(\ln\sqrt{h})_{,u}\,r$ term (all the coefficients of the powers of $r$ appearing in $H$ are indeed constants) and represents static black holes. Again it follows that the Weyl type is D(bd) (possibly, D(bcd)) {and $\bl$ is also a double WAND.} The Maxwell field is given by \eqref{F_generic} with $2p=n$. Here \eqref{peeling_n/2} clearly reduces to ${\bF=\mbox{\boldmath{$D$}}\,r^{-\frac{n}{2}}}$. These $2p=n$ solutions with $\bF$ of type D were previously studied in \cite{BarCalCha12}. Two examples with $n=8$, $p=4$ are given by \eqref{metric_n=8_p=4_II} and \eqref{metric_n=8_p=4_II_2} with $f_{ijk}(u)=0$ (and thus $\mu=\mbox{const}$), see \cite{BarCalCha12} for others.

\subsection{Type N Maxwell field}

\label{subsec_null_Max}

The field \eqref{F_2p=n} is of type N iff
\be
 \e_{i_1\ldots i_{p-2}}=0 , \qquad \bb_{i_1\ldots i_{p}}=0 ,
\ee
with $\bk$ being {the unique} aligned null direction, so that
\beqn
 \bF= & & \frac{1}{\left(\frac{n}{2}-1\right)!}\,{f}_{i_1\ldots i_{p-1}}\d u\wedge\d x^{i_1}\wedge\ldots\wedge\d x^{i_{p-1}} , \label{F_2p=n_N}
\eeqn
and the peeling \eqref{peeling_n/2} becomes simply $\bF=\mbox{\boldmath{$N$}}r^{1-\frac{n}{2}}$.

In this case {eqs.~\eqref{maxw_2p_1}--\eqref{mu_u_2p=n} reduce to}
\beqn
    & & {f}_{[i_2 \ldots i_{p},i_1]}=0 , \qquad {(\sqrt{h}\, {f}^{k i_1 \ldots i_{p-2}})_{,k}=0 } , \label{maxw_2p_2_N} \\
		& & \mu_{,i} = 0 , \label{mu_i_2p=n_N} \\
		& & (n-2)\mu_{,u}=-(n-1)\mu(\ln\sqrt{h})_{,u}{-}2\b \F^2 \qquad (n>4)  \label{mu_u_2p=n_N} .
\eeqn
Eqs.~\eqref{maxw_2p_2_N} mean that ${f}_{i_1 \ldots i_{p-1}}(u,x)$  effectively defines a Maxwell $\left(\frac{n}{2}-1\right)$-form in the $(n-2)$-dimensional transverse space  (i.e., it is harmonic).

The line-element is \eqref{ds_generic}, where $h_{ij}(u,x)=h^{1/(n-2)}(u,x)\,\gamma_{ij}(x)$ is Einstein and with $H(u,r,x)$ given by
\be 2H=K{+}\frac{2(\ln\sqrt{h})_{,u}}{n-2}\,r-\frac{2\Lambda}{(n-1)(n-2)}\,r^2-\frac{\mu(u)}{r^{n-3}} . \label{grr_2p=n_N}
\ee
The term $(\ln\sqrt{h})_{,u}$ can be reexpressed in terms of $\mu$ and $\F^2$ using \eqref{mu_u_2p=n_N}, if desired. Note that by \eqref{mu_u_2p=n_N} necessarily $\mu\neq 0$.\footnote{It is worth observing that this is not true in {4D}, due precisely to the additional term proportional to $\triangle{\cal R}$ entering \eqref{mu_u_2p=n_N} when $n=4$. Thanks to this extra term, {4D} solutions of Petrov type III (with $\mu=0$ {and a null Maxwell field}) are thus possible, see section~28.2.2 of \cite{Stephanibook} and references therein.} {Since ${f}_{i_1 \ldots i_{p-1}}$ is harmonic, the metric $h_{ij}$ cannot be of positive constant curvature \cite{Bochner48,YanBocbook}.}

Here the energy-momentum tensor possesses only the (b.w. -2) component describing a flux of radiation {along $\bk$}
\be
  8\pi T_{\mu\nu} l^\mu l^\nu=8\pi T_{uu}=\b r^{2-n}\F^2 ,
\ee
and thus the Maxwell form acts as {an aligned} {\em pure radiation} field. Therefore these special Robinson--Trautman metrics are contained in the pure radiation family of solutions studied in \cite{PodOrt06} (but the corresponding Maxwell equations were not considered there). In particular, from appendix~A of \cite{PodOrt06} it immediately follows that the Weyl type is D(bd) (D(bcd) if $h_{ij}$ is of constant curvature), with \eqref{nulldirections} being the two multiple WANDs {and
\be
 \Phi=-(n-2)(n-3)\frac{\mu(u)}{2r^{n-1}}\neq 0  , \qquad \tilde\Phi_{\hat i\hat j\hat k\hat l}=r^{-2}{\cal C}_{\bar i\bar j\bar k\bar l} ,
\ee
as can also be seen explicitly from \eqref{weyl0_text}--\eqref{weyl-2_text} with $\E^2=0=\B^2$.}

\subsubsection{A special subcase: solutions with a factorized $h(u,x)=U(u)X(x)$}

In the special case of a factorized $h(u,x)=U(u)X(x)$, one can set $U(u)=1$ by a coordinate transformation, and thus obtain a special subclass of solutions with metric \eqref{ds_generic} with $h_{ij}=h_{ij}(x)$ and $H(u,r)$ given by
\beqn
	 & & 2H=K-\frac{2\Lambda}{(n-1)(n-2)}\,r^2-\frac{\mu(u)}{r^{n-3}} , \\
	 & & \mu(u)=\mu_0{-}\frac{2\b}{n-2}\int {\F^2} \,\d u .
\eeqn

Here necessarily $\mu_{,u}\neq 0$. The Maxwell field is given by \eqref{F_2p=n_N}, where ${f}_{i_1\ldots i_{p-1}}$ must satisfy \eqref{maxw_2p_2_N} and, {by \eqref{mu_u_2p=n_N}, also}
\be
  (\F^2)_{,i}=0 . \label{2p=n_const_inv}
\ee
{These solutions will in general describe formation of black holes in the presence of electromagnetic radiation {with non-zero expansion}, with a monotonically increasing (or decreasing, {according to the choice of time-orientation of $\bk$}) mass parameter $\mu$.}

A simple solution can be obtained by taking the transverse metric to be flat, i.e., $h_{ij}=\delta_{ij}$ and ${f}_{i_1\ldots i_{p-1}}=f_{i_1\ldots i_{p-1}}(u)$ (obviously with such a choice \eqref{maxw_2p_2_N} and \eqref{2p=n_const_inv} are identically satisfied). To be specific, we give the following explicit example for $n=6$.

\paragraph{Example ($n=6$, $p=3$)} The metric and the Maxwell field are
\beqn
 & & h_{ij}=\delta_{ij} , \qquad 2H=-\frac{\Lambda}{10}r^2-\frac{\mu(u)}{r^3} , \qquad \mu(u)=\mu_0{-}\frac{\b}{2}\int\F^2\d u , \nonumber \\
 & & F_{uij}=f_{ij}(u) . \label{6D_null}
\eeqn
For $\Lambda<0$ this spacetime represents the formation of asymptotically locally AdS black holes with electromagnetic radiation. By a rotation, one can always simplify the Maxwell field so as to only have non-zero components $F_{u12}=f_{12}(u)$, $F_{u34}=f_{34}(u)$, in which case $\bF$ is {self-dual when $f_{34}(u)=-f_{12}(u)$ (or anti-self-dual when $f_{34}(u)=+f_{12}(u)$), cf. section~\ref{subsec_2p=n_summary}}. Solution~\eqref{6D_null} is an extension of a solution given in {4D} (for $\Lambda=0$) in \cite{RobTra62} (also reproduced in eq.~(28.43) of \cite{Stephanibook}) {and recently discussed in \cite{Senovilla14}} {(see also footnote~\ref{footn_RTnull})}. {It should be observed that for $n=6$ this example in fact comprises all the possible Robinson--Trautman solutions with a Maxwell field of type N {\em if} the transverse space is assumed to be of constant curvature (this follows from \eqref{maxw_2p_2_N} and \eqref{2p=n_const_inv} after using coordinates adapted to the constant curvature space, e.g., those employed in \cite{PodOrt06} --- it also follows that the (constant) curvature must necessarily be zero). A similar example with $n=8$, $p=4$ is given by \eqref{metric_n=8_p=4_II} with $\E=0=\B$. An $n=6$ example where, instead, $h_{ij}$ is not of constant curvature follows.}

\paragraph{Example  ($n=6$, $p=3$)} As another example in {6D} one can take $h_{ij}$ to be a direct product of two $S^2$ or $H^2$, namely
\beqn
 & & h_{ij}\d x^i\d x^j=\left[1-3 K  x_1^2\right]^{-1}\d x_1^2+\left[1-3 K  x_1^2\right]\d x_2^2+\left[1-3 K  x_3^2\right]^{-1}\d x_3^2+\left[1-3 K  x_3^2\right]\d x_4^2 , \nonumber \\
 & & 2H= K -\frac{\Lambda}{10}r^2-\frac{\mu(u)}{r^3} , \qquad  K =\pm 1 , \qquad \mu(u)=\mu_0{-}\b\int\left(f_{12}^2(u)+f_{34}^2(u)\right)\d u , \label{6D_null_K} \\
 & & F_{u12}=f_{12}(u), \qquad F_{u34}=f_{34}(u) .  \nonumber
\eeqn
{As above, (anti-)self-duality holds iff $f_{34}(u)=\mp f_{12}(u)$.} For $K =0$ this solutions reduces to \eqref{6D_null} (up to a space rotation).

\section*{Acknowledgments}
%\acknowledgments

M.O. and M.\v Z. have been supported by the Albert Einstein Center for Gravitation and Astrophysics, Czech Science Foundation GA\v{C}R~14-37086G. M.O. has also been supported by research plan {RVO: 67985840}. J.P. has been supported by the research grant GA\v{C}R~P203/12/0118.

\renewcommand{\thesection}{\Alph{section}}
\setcounter{section}{0}

\renewcommand{\theequation}{{\thesection}\arabic{equation}}

\section{Robinson-Trautman spacetimes with an aligned Ricci tensor of type II}
\setcounter{equation}{0}

\label{app_summaryRT}

For the purposes of the present paper and for possible future reference it is useful to summarize some of the results of {\cite{PodOrt06,OrtPodZof08,SvaPod14,PodSva14}} in the present appendix. Note that we restrict here to the Robinson-Trautman spacetimes in which the Ricci tensor is of type II aligned with the privileged vector field $\bk$. This includes vacuum solutions as well as solutions with aligned matter content {(as we assume in the main text).}

\subsection{General metric}

The line-element and its Weyl type are specified by the following theorem.

\begin{theorem}
\label{theor_RT}

If a $n$-dimensional spacetime {($n\ge 4$)} admits a non-twisting, non-shearing, expanding geodesic null vector field~{\boldmath $k$} and the Ricci tensor is of aligned type II, adapted coordinates $(u,r,x^1,\ldots,x^{n-2})$ can be chosen such that \cite{PodOrt06}
\beqn
  & & \d s^2=r^{2}h_{ij}\left(\d x^i+ W^{i}\d u\right)\left(\d x^j+ W^{j}\d u\right){-}2\,\d u\d r-2H\d u^2 , \label{geo_metric} \\
	& & h_{ij}=h_{ij}(u,x) , \qquad W^{i}={\alpha}^i(u,x)+r^{1-n}{\beta}^i(u,x) , \label{h_W} \\
	& & \bk=\pa_r , \qquad \theta=1/r ,
\eeqn
where $H$ is an arbitrary function of all coordinates. $\bk$ is automatically a WAND, such that the Weyl tensor is in general of aligned type I(b). It is a multiple WAND iff ${\beta}^i=0$ \cite{PodSva14}, in which case the Weyl tensor is of aligned type II(d) (or more special). {When ${\beta}^i=0$, one can locally set $W^{i}=0$ (after a coordinate transformation giving ${\alpha}^i=0$) \cite{PodOrt06}.} {The Weyl type further specializes to II(bd) iff $h_{ij}$ is an Einstein metric (with still $W^{i}=0$) \cite{PodSva14}.}

\end{theorem}

The vector field {\boldmath $k$} is the generator of null hypersurfaces $u=\,$const such that $k_\mu\d x^\mu={-}\d u$, $r$ is an affine parameter along $\bk$, $\theta$ is its expansion scalar, and $x \equiv (x^i) \equiv (x^1, \ldots, x^{n-2})$ are spatial coordinates on a ``transverse'' $(n-2)$-dimensional Riemannian manifold. We observe that the condition that the Ricci tensor is doubly aligned with $\bk$ can be expressed in a frame-independent form as $R_{\mu\nu}k^\nu=\a k_\mu$, which in the coordinates of \eqref{geo_metric} means $R_{rr}=0=R_{ri}$. For certain calculations it may be useful to observe that $2H=g^{rr}=-g_{uu}$ and $W^i=g^{ri}$ (such that $W^i=0\Leftrightarrow g^{ri}=0\Leftrightarrow g_{ui}=0$).

The first part of the theorem was proven in sections 3.1 and 3.2 of \cite{PodOrt06}. The results on Weyl alignment follow immediately from eqs. (13)--(15), {(17) and (19) (with (24), (25))} of \cite{PodSva14} together with \eqref{geo_metric}.\footnote{Recall that the Weyl type I(b) in fact characterizes {\em all} Robinson--Trautman geometries, independently of any assumptions on the Ricci tensor \cite{PodSva14}. This can also be seen from the Ricci identities (11g) and (11k) (and their trace) of \cite{OrtPraPra07} using the fact that $\bk$ is geodesic, shearfree and twistfree.} In vacuum or with aligned pure radiation (i.e., the only non-zero energy-momentum tensor component being $T_{uu}$) necessarily ${\beta}^i=0$ {and $h_{ij}$ is Einstein} for any $n\ge 4$, and the Weyl tensor further specializes to type D(bd) {(possibly, D(bcd) or D(acd))} if $n>4$ \cite{PodOrt06} (the latter result was also rederived in \cite{PodSva14} without the coordinate choice ${\alpha}^i=0$), {the type O being possible only in the trivial case of constant curvature spacetimes.}  In \cite{SvaPod14} it was shown that for any $n\ge 4$ Robinson--Trautman spacetimes cannot support aligned gyratonic matter (i.e., an energy-momentum tensor with {\em both} $T_{uu}$ and $T_{ui}$ --- and no other components --- being non-zero). Let us finally observe that if one relaxes the assumptions of the theorem by requiring only the aligned Ricci type I (i.e., $R_{rr}=0$), one obtains the same form of the metric, except that the $W^i(u,r,x)$ are arbitrary functions  \cite{PodOrt06} (the Weyl type remains I(b) \cite{PodSva14}).

\subsection{Ricci tensor components}

\label{subapp_ricci}

The Ricci tensor component $R_{ij}$ for the general metric \eqref{geo_metric}, \eqref{h_W} was given explicitly in eq.~(A.1) of \cite{OrtPodZof08} and reads
\begin{eqnarray}
 && R_{ij}\>=\> {\mathcal R_{ij}}-r^{4-n}\left(r^{n-3}2H\right)_{,r}h_{ij}-r^{2(2-n)}\frac{(n-1)^2}{2}h_{ik}h_{jl}  {\beta}^{k}{\beta}^{l} \nonumber\\
 && {\qquad}-r\left[\frac{n-2}{2}\Big(2h_{k(i}{{\alpha}^{k}}_{,j)}+{\alpha}^{k}h_{ij,k}{-}h_{ij,u}\Big)
    +\left({\alpha}^{k}_{\;,k}+{\alpha}^{k}(\ln\sqrt{h})_{,k}{-}(\ln\sqrt{h})_{,u}\right)h_{ij}\right]\nonumber\\
 && {\qquad}+r^{2-n}\left[\frac{1}{2}\left(2h_{k(i}{{\beta}^{k}}_{,j)}+{\beta}^{k}h_{ij,k}\right)
    -\left({\beta}^{k}_{\;,k}+{\beta}^{k}(\ln\sqrt{h})_{,k}\right)h_{ij}\right] , \label{Rij_general}
\end{eqnarray}
where ${\mathcal R_{ij}}$ is the Ricci tensor associated with the spatial metric $h_{ij}$ (as defined in section~\ref{subsec_Rur}), and a partial derivative w.r.t. (e.g.) $x^j$ is simply denoted by a comma followed by $j$. {For the purposes of the present paper we need the remaining Ricci components only in the special case $W^i=0$, which we now present (but see \cite{SvaPod14,PodSva14} for the Ricci tensor components of the most general Robinson--Trautman metric, i.e., without not even enforcing $R_{rr}=0=R_{ri}$).}

\subsubsection{Case $W^i=0$ ($\bk$ is a multiple WAND --- Weyl type II(bd))}

{\em If one further assumes $W^i=0$} in \eqref{geo_metric} ({i.e., $\bk$ is a multiple WAND,} as we indeed find in section~\ref{subsec_Rur}), {eq.~\eqref{Rij_general} and} the remaining Ricci components, given in eqs.~(26), (27) and (31) of \cite{PodOrt06}, reduce to
\beqn
 R_{ij}\> & =& \> {{\mathcal R_{ij}}-r^{4-n}\left(r^{n-3}2H\right)_{,r}h_{ij}{+}r\left[\frac{n-2}{2}h_{ij,u}+(\ln\sqrt{h})_{,u}h_{ij}\right]} , \label{Rij_general_W=0} \\
 R_{ur} & = & r^{2-n}\left(r^{n-2}H_{,r}\right)_{,r}-r^{-1}(\ln\sqrt{h})_{,u}   , \label{Rur_general} \\
 R_{ui}&=& r^{4-n}\left(r^{n-4}H_{,i}\right)_{,r}
  + \frac{1}{2}\left(h^{jk}h_{ik,u}\right)_{,j} \nonumber\\
 && +\frac{1}{2}h^{jk}h_{ik,u}(\ln\sqrt{h})_{,j} -\frac{1}{4}h^{jk}h^{lm}h_{kl,u}h_{jm,i}-(\ln\sqrt{h})_{,ui} ,    \label{Rui_general} \\
 R_{uu}&= & 2HR_{ur} {-}r^2(r^{-2}H)_{,r}(\ln\sqrt{h})_{,u}
{+}(n-2)r^{-1}H_{,u} \nonumber \\
&& +r^{-2}\triangle H -(\ln\sqrt{h})_{,uu}-\frac{1}{4}h^{il}h^{jk}\,h_{ij,u}h_{kl,u}      , \label{Ruu_general}
\eeqn
where in the last expression the first quantity on the r.h.s. has been written in terms of \eqref{Rur_general} for brevity and for convenience in the calculations of section~\ref{subsubsec_Ruu_gener}, and $\triangle\equiv\frac{1}{\sqrt{h}}\pa_j(\sqrt{h}h^{ij}\pa_i)$ is the Laplace operator in the $(n-2)$-dimensional space with metric $h_{ij}$. Note that $2(n-2)R_{ur}={-}r^{-1}\left(h^{ij}R_{ij}\right)_{,r}$. {For general purposes it is useful to observe that the component $R_{ui}$ can be rewritten more compactly as $R_{ui} = r^{4-n}(r^{n-4}H_{,i})_{,r}+\frac{1}{2}h^{kl}(h_{ki,u||l}-h_{kl,u||i})$ \cite{SvaPod14}, where the lower double bar $_{||}$ denotes a covariant derivative w.r.t. $h_{ij}$ --- but in the computations of the present paper the form \eqref{Rui_general} can be used more readily.}

\subsection{Weyl tensor components in the case $R_{ij}\propto h_{ij}$ ($\Rightarrow W^i=0$)}

\label{app_Weyl}

For the purposes of the present paper, we further restrict here to the case when the condition $R_{ij}\propto h_{ij}$ holds for the metric \eqref{geo_metric} (this is also true in various other cases of physical interest, e.g., in vacuum, or whenever the traceless Ricci tensor is of aligned type III or more special), while we refer to \cite{PodSva14} for the Weyl tensor components of the most general Robinson--Trautman metric. Similarly as in section~\ref{subsec_Rur}, together with \eqref{Rij_general} this restriction gives $\beta^i=0$, and thus $W^i=0$ (theorem~\ref{theor_RT}). By \eqref{Rij_general_W=0} we further have
\be
  \R_{ij}=\frac{\R}{n-2}h_{ij} , \qquad h_{ij,u} =\frac{2(\ln\sqrt{h})_{,u}}{n-2}h_{ij} ,
	\label{Ricci_restric}
\ee
so that, in particular, $h_{ij}$ is Einstein and therefore $(n-4)\R_{,i}=0$. (Note that, indeed, these conditions are obtained in section~\ref{subsec_Rur} as a consequence of the Einstein equations --- except in the cases $p=1,n-1$, {for which see appendix~\ref{app_limiting}}.) Using \eqref{Ricci_restric} one can further prove the identity (cf. eq.~(144) of \cite{PodSva14})
\be
 (n-4)\left[[(\ln\sqrt{h})_{,u}]_{||ij}-\frac{\triangle[(\ln\sqrt{h})_{,u}]}{n-2}h_{ij}\right] =0 ,
\label{identity_||}
\ee
{needed in the following.}

We choose a frame
\be
 \bk=\pa_r , \qquad \bl={-}\pa_u+H\pa_r , \qquad \bbm_{(\hat i)}=\frac{1}{r}m_{(\hat i)}^j\pa_j ,
 \label{frame}
\ee
where the functions $m_{(\hat i)}^j$ do not depend on $r$. The spacelike vectors $\bbm_{(\hat i)}$ span the transverse space of constant $u$ and $r$, and obviously $\tbbm_{(\tilde i)}\equiv r \bbm_{(\hat i)}=m_{(\hat i)}^j\pa_j $ defines an orthonormal frame for the metric $h_{ij}$. {Before proceeding, it is useful to observe that since $\bk$ and $\bl$ span the 2-space $(u,r)$ we have the relation $\tau_{i}-\tau'_{i}=0$ (in the notation of \cite{Durkeeetal10,OrtPraPra13rev} for the Ricci rotation coefficients). Using \eqref{ds_generic}, it immediately follows that $\bl$ is parallelly transported along $\bk$, i.e.,  $\tau'_{i}=0$, so that also $\tau_{i}=0$. There cannot be other null directions $\bl'$ with these properties (as follows from the transformation properties of $\tau_{i}$ and $\tau'_{i}$ under null rotations \cite{OrtPraPra07,Durkeeetal10}). It is also easy to see that {\em $\bl$ is geodesic iff $H_{,i}=0$} \cite{PraPraOrt07}. In fact, not only $\bl$  but the full frame~\eqref{frame} is parallelly transported along $\bk$.

Recall that, thanks to $W^i=0$ and the first of \eqref{Ricci_restric}, the Weyl type here is II(bd) or more special (Theorem~\ref{theor_RT}). In the frame~\eqref{frame} one further finds $\tilde\Psi'_{\hat i\hat j \hat k}=0$, so that all the non-zero components are determined by (using eqs.~(16), (18), (20) and (22) of \cite{PodSva14}, or eqs.~(A.1) of \cite{PodOrt06})
\beqn
 & & \Phi=\frac{n-3}{2(n-1)}\left\{r^2\left[r^{-2}\left(2H-\frac{\R}{(n-2)(n-3)}\right)\right]_{,r}\right\}_{,r} , \qquad \tilde\Phi_{\hat i\hat j\hat k\hat l}=r^{-2}{\cal C}_{\bar i\bar j\bar k\bar l} , \label{weyl0} \\
 & & \Psi'_{\hat i}=rm_{(\hat i)}^k\frac{n-3}{n-2}\left[r^{-2}\left(H{-}\frac{(\ln\sqrt{h})_{,u}}{n-2}\,r\right)\right]_{,kr} , \label{weyl-1} \\
 & & \Omega'_{\hat i\hat j}=\frac{1}{r^2}m_{(\hat i)}^k m_{(\hat j)}^l\left(H_{||kl}-\frac{\triangle H}{n-2}h_{kl}\right) , \label{weyl-2}
\eeqn
where ${\cal C}_{\bar i\bar j\bar k\bar l}$ are the frame components of the Weyl tensor associated with $h_{ij}$ (in the corresponding frame $\tbbm_{(\bar i)}$). The above components are ordered by b.w. and expressed in the notation of \cite{Durkeeetal10,OrtPraPra13rev} (cf. table~2 of \cite{OrtPraPra13rev}). We observe that the alternative notation of \cite{PodSva14} corresponds to $\Psi_{2S}=-\Phi$,  $\tilde\Psi_{2^{ijkl}}=\tilde\Phi_{\hat i\hat j\hat k\hat l}$,  $\Psi_{3T^{i}}=\Psi'_{\hat i}$ and $\Psi_{4^{ij}}=\Omega'_{\hat i\hat j}$.

\subsubsection{Conditions for the Weyl types II(abd), II(bcd) and III(b)}

\label{subsubsec_weyl_III}

As noticed above, here the Weyl type is generically II(bd). It becomes II(abd) iff $\Phi=0$, i.e. (using the first of \eqref{weyl0}), when
\be
 2H=\frac{\R}{(n-2)(n-3)}+c_1(u,x)r+c_2(u,x)r^2 ,
\label{II(abd)}
\ee
where $c_1$ and $c_2$ are integration functions independent of $r$ (recall that $\R$ is also independent of $r$ and $(n-4)\R_{,i}=0$, as noticed above).

On the other hand, the type becomes II(bcd) iff $\tilde\Phi_{\hat i\hat j\hat k\hat l}=0$, namely (using the second of \eqref{weyl0}) for
\be
 {\cal C}_{\bar i\bar j\bar k\bar l}=0 ,
\label{II(bcd)}
\ee
which (with the first of \eqref{Ricci_restric}) means that $h_{ij}$ is a constant curvature metric.

The Weyl type further specializes to III(b) (recall that here we have $\tilde\Psi'_{\hat i\hat j \hat k}=0$ identically) iff \eqref{II(abd)} and \eqref{II(bcd)} hold simultaneously.

\subsubsection{Conditions for the Weyl type N}

\label{subsubsec_weyl_N}

The Weyl type becomes N when $\Phi=0$, $\tilde\Phi_{\hat i\hat j\hat k\hat l}=0$ and $\Psi'_{\hat i}=0$ which gives (using \eqref{weyl-1} and \eqref{II(abd)})
\be
	 2H=\frac{{\cal R}}{(n-2)(n-3)}{+}\frac{2(\ln\sqrt{h})_{,u}}{n-2}\,r+c_2(u,x)r^2 ,
 \label{N}
\ee
which must hold together with  and \eqref{II(bcd)}. {In this case, for $n>4$ the only non-zero Weyl components are (using also \eqref{identity_||})
\be
 \Omega'_{\hat i\hat j}=\frac{1}{2}m_{(\hat i)}^k m_{(\hat j)}^l\left(c_{2||kl}-\frac{\triangle c_2}{n-2}h_{kl}\right) \qquad (n>4) .
\ee
See \cite{Stephanibook} for $n=4$.}

\subsubsection{Conditions for $\bl$ to be a multiple WAND ($\Rightarrow$ Weyl type D(bd))}

\label{app_l_mWAND}

It is clear from \eqref{weyl-1} and \eqref{weyl-2} that, in general, the vector field $\bl$ of \eqref{frame} is not a WAND (not even a single one).  The conditions for $\bl$ being a multiple WAND read
\beqn
	 & & 2H=d_1(u,r){+}\frac{2(\ln\sqrt{h})_{,u}}{n-2}\,r+c_2(u,x)r^2 , \label{n_mWAND} \\
	 & & c_{2||ij}=\frac{\triangle c_2}{n-2}h_{ij} , \qquad [(\ln\sqrt{h})_{,u}]_{||ij}=\frac{\triangle[(\ln\sqrt{h})_{,u}]}{n-2}h_{ij} . \label{n_mWAND_2}
\eeqn
{For $n>4$ the latter of these is identically satisfied thanks to \eqref{identity_||}. When all these conditions are met the Weyl type becomes D(bd). This happens, for example, in the special case $c_{2,i}=0$.}

\subsubsection{Conditions for the Weyl type D(bd) when $\bl$ is {\em not} a multiple WAND}

\label{app_type_D(bd)}

Even when the vector field $\bl$ of \eqref{frame} is not a multiple WAND (i.e., \eqref{n_mWAND}, \eqref{n_mWAND_2} are not {simultaneously} satisfied) the Weyl type can still be D(bd) provided a second multiple WAND (in addition to $\bk$, and different from $\bl$) exists. In order to find conditions for this to happen, it is necessary to perform a null rotation about $\bk$, i.e.,
\be
 \bk\mapsto\bk, \qquad \bl\mapsto\bl+z_{\hat i}\bbm_{\hat i} -\frac{1}{2} z_{\hat i}z^{\hat i}\bk , \qquad \bbm_{\hat i} \mapsto \bbm_{\hat i} -z_{\hat i}\bk ,
    \label{nullrot}
\ee
such that, in the transformed frame, $\Psi'_{\hat i}\mapsto 0$, $\tilde\Psi'_{\hat i\hat j \hat k}\mapsto0$ and $\Omega'_{\hat i\hat j}\mapsto0$. The transformation laws under \eqref{nullrot} for the negative b.w. Weyl components are given by eqs.~(2.33)--(2.35) of \cite{Durkeeetal10} (while non-negative b.w. components are unchanged under \eqref{nullrot} since $\bk$ is a multiple WAND). Using the fact that the Weyl type is II(bd), the condition $\Psi'_{\hat i}\mapsto 0$ uniquely fixes the parameters $z_{\hat i}$ by
\be
 \Phi z_{\hat i}=\frac{n-2}{n-1}\Psi'_{\hat i} , \label{z_cond}
\ee
except in the case $\Phi=0=\Psi'_{\hat i}$, for which the $z_{\hat i}$ remain arbitrary.

Next, requiring $\tilde\Psi'_{\hat i\hat j \hat k}\mapsto0$ and using \eqref{z_cond} further gives (also in the case $\Phi=0=\Psi'_{\hat i}$)
\be
 \tilde\Psi'_{\hat i\hat j \hat k}=-\tilde\Phi_{\hat l\hat i\hat j\hat k}z_{\hat l} .
 \label{z_cond_2}
\ee

Finally, imposing $\Omega'_{\hat i\hat j}\mapsto0$ gives

\be
 \Omega'_{\hat i\hat j}= \frac{2(n-2)}{n-3}\left(z_{\hat (j}\Psi'_{\hat i)}-\frac{(z_{\hat k}\Psi'_{\hat k})}{n-2}\delta_{\hat i\hat j}\right)+2z_{\hat k}\tilde\Psi'_{(\hat i\hat j) \hat k}-\frac{n-1}{n-3}\Phi\left(z_{\hat i}z_{\hat j}-\frac{z^2}{n-2}\delta_{\hat i\hat j}\right)-z_{\hat k}z_{\hat l}\tilde\Phi_{\hat k\hat i\hat l\hat j} ,
 \label{z_cond_3_gen}
\ee
which after using \eqref{z_cond}, \eqref{z_cond_2} (and multiplying by $\Phi$) gives a constraint among the Weyl tensor components\footnote{In passing, let us observe that \eqref{z_cond}, \eqref{z_cond_2} and \eqref{z_cond_3} give the necessary and sufficient conditions under which {\em any} Weyl tensor of type II(bd) admits a second multiple WAND (so being in fact of type D(bd)), since no special features of the Robinson-Trautman class have been used in their derivation. For the same reason, \eqref{z_cond_3_gen} alone determines the conditions under which a Weyl tensor of type II(bd) admits a single WAND (in addition to a multiple one).
This implies that a Weyl tensor of {\em proper} type D(bd) or D(bcd) admits {\em precisely two} WANDs (necessarily double) since \eqref{z_cond_3_gen} has no non-zero solution if $\Omega'_{\hat i\hat j}=\Psi'_{\hat i\hat j\hat k}=\Psi'_{\hat i}=0$, $\Phi\neq0$ (as easily seen after contraction with $z_{\hat j}$). However, Weyl tensors of type D(abd) may admit an infinity of multiple WANDs, and this occurs precisely when the equation $\tilde\Phi_{\hat l\hat i\hat j\hat k}z_{\hat l}=0$ (cf.~\eqref{z_cond_2}) admits a solution (in this case \eqref{z_cond} and \eqref{z_cond_3} are satisfied trivially) --- see, e.g., \cite{OrtPraPra13,OrtPraPra13rev} for examples of such spacetimes. Similarly, an infinity of {\em single} WANDs exists for type D(abd) if  the weaker condition $z_{\hat k}z_{\hat l}\tilde\Phi_{\hat k\hat i\hat l\hat j}=0$ is satisfied. We observe that L. Wylleman has obtained more general results on the structure of multiple WANDs for any Weyl type D (private communication), some of which are mentioned in \cite{HerOrtWyl13,OrtPraPra13rev}.\label{footn_WANDs}}
\be
 \Phi\Omega'_{\hat i\hat j}=\frac{n-2}{n-1}\left[\frac{n-2}{n-3}\left(\Psi'_{\hat i}\Psi'_{\hat j} -\frac{(\Psi'_{\hat k}\Psi'_{\hat k})}{n-2}\delta_{\hat i\hat j}\right)+\tilde\Psi'_{(\hat i\hat j) \hat k}\Psi'_{\hat k}\right] . \label{z_cond_3}
\ee

Recalling that for the metric \eqref{geo_metric} with $W^i=0$ and \eqref{Ricci_restric} one has $\tilde\Psi'_{\hat i\hat j \hat k}=0$, and using \eqref{weyl0}, eq.~\eqref{z_cond_2} reduces to
\be
 {\cal C}_{\bar l\bar i\bar j\bar k}z_{\hat l}=0 .
\ee
This is a restriction on the Weyl tensor of $h_{ij}$ that will not be true for a {\em generic} Einstein metric $h_{ij}$, therefore we can conclude that {\em generically the metric \eqref{geo_metric} with $W^i=0$ and \eqref{Ricci_restric} is of genuine type II(bd)}. However, the ``genericity'' conditions may be violated if a special choice of $H$ (e.g., in vacuum \cite{PodOrt06}) or of $h_{ij}$ is made in \eqref{geo_metric}, in which case the Weyl types D(bd), D(bcd), and D(abd) are all possible (see, e.g., \cite{PodOrt06,OrtPodZof08,OrtPraPra13rev,PodSva14}). The specific form of \eqref{z_cond_3} for spacetimes \eqref{geo_metric} with $W^i=0$ and \eqref{Ricci_restric} can be obtained by substituting \eqref{weyl0}--\eqref{weyl-2} and $\tilde\Psi'_{\hat i\hat j \hat k}=0$ into \eqref{z_cond_3}.

\section{Comments on the (Einstein) constraints \eqref{eqE} and \eqref{eqB} {($2\le p\le n-2$)}}

\setcounter{equation}{0}

\label{app_comments}

The b.w.~0 components of $\bF$ can be divided into an electric and a magnetic part described, respectively, by $\e_{i_1\ldots i_{p-2}}$ and $\bb_{i_1\ldots i_p}$ (cf. \eqref{Max0_r}, \eqref{Max0_ur}). The latter live in the transverse geometry of $h_{ij}$ and must obey the constraints \eqref{eqE} and \eqref{eqB} (replaced by \eqref{eqEB} if $2p=n$). {In this appendix we discuss some general consequences of those constraints.}

Recall that the form $\e_{i_1\ldots i_{p-2}}$ is defined for $2\le p\le n$ while $\bb_{i_1\ldots i_p}$ for $0\le p\le n-2$. The cases  $p=0,n$ are trivial (footnote~\ref{footn_trivial}) while $p=1,n-1$ require a special discussion (appendix~\ref{app_limiting}), so here we restrict for both $\e_{i_1\ldots i_{p-2}}$ and $\bb_{i_1\ldots i_p}$ to the ranks $2\le p\le n-2$. Let us also observe that the equations obeyed by $\e_{i_1\ldots i_{p-2}}$ and $\bb_{i_1\ldots i_p}$ (\eqref{eqE} and \eqref{eqB}) are identical, so the algebraic constraints derived from them which apply to $\bb_{i_1\ldots i_p}$ for a certain $p$ will also automatically apply to $\e_{i_1\ldots i_{p'-2}}$ for $p'=p+2$ (and vice versa). Additionally, by duality constraints on $\bb_{i_1\ldots i_p}$ for a certain $p$ will also apply on $\e_{i_1\ldots i_{p'-2}}$ for $p'=n-p$  (and vice versa). In the magnetic and electric cases the following ranks $p$ are worth mentioning.

\subsection{Magnetic fields}

\label{subsec_magn}

\begin{itemize}

	\item $p=2$: this is the case studied in \cite{OrtPodZof08} and $\bb_{ij}=0$ if $n$ is odd, so that $n$ must be even (thus, in particular, the $\ln r$ term in $2H$ (eq.~\eqref{H_log-}) vanishes for $n=5$ \cite{OrtPodZof08}). If $\bF$ is assumed to be regular (and non-zero) on the transverse space, then Maxwell's equations \eqref{diverg_elec}, \eqref{diverg_magn} for $\bb_{ij}$ further imply \cite{OrtPodZof08} that $h_{ij}$ must be almost-K\"ahler (almost-Hermitian if $n=6$). See eq.~(66) of \cite{OrtPodZof08} for an example.

	\item $p=n-4$: this is dual to the {electric $p=4$ case} discussed in section~\ref{subsec_elec} so that again $n$ must be even and $h_{ij}$ almost-K\"ahler (almost-Hermitian if $n=6$).
	
	\item $p=n-3$: it is easy to see that in this case eq.~\eqref{eqB} is impossible (unless $\bb_{i_1\ldots i_{n-3}}=0$). Using also the previous observations one sees, e.g., that for $n=5$ only $p=3$ is permitted, for $n=6$ only $p=2,4$, for $n=7$ only $p=5$, etc. This implies, in particular, that the $\B^2r^{3-n}\ln r$ term in $2H$ (eq.~\eqref{H_log-}) can occur only for $p=(n+1)/2\ge 4$ with $n\ge 9$ ($n$ odd).

	\item $p=n-2$: $F_{i_1\ldots i_p}$ is proportional to the $h$-volume element. For example, this includes the magnetic counterpart of the electrovac Schwarzschild-Tangherlini black hole \cite{Tangherlini63}.

	\item $p=\frac{n}{2}-1$ ($n$ even): in this case {if $h_{ij}$ is a direct product of two ($\frac{n}{2}-1$)-dimensional spaces then} a possible magnetic field can be written as a sum of their volume elements, see also \cite{BarCalCha12}; for even $p$ it is $h$-self-dual. 	 This case is also special in that the magnetic field does not contribute to $T_{ij}$. In particular, this includes the case $n=6$, $p=2$ (again eq.~(66) of \cite{OrtPodZof08} --- with $D=6$ --- gives an example). See {\eqref{el_n=8_p=5}} for an example in 8D (after dualization).

 \item $p=md$, $n-2=Nd$: when $p$ and $n-2$ are both multiples of the same integer $d\ge2$ ($m$ and $N$ are also positive integers, with $m\le N$), eq.~(3.18) of \cite{BarCalCha12} gives a method of constructing $F_{i_1\ldots i_p}$ that works when $h_{ij}$ is a direct product of $N=(n-2)/d$ $d$-dimensional Einstein spaces (in particular they are all flat if $\R=0$). (For $N=2$, $m=1$ this reduces to the case $p=\frac{n}{2}-1$ discussed above.) For example: $n=8$, $p=3$, $d=3$, $N=2$, $m=1$ (note that here $p=\frac{n}{2}-1$); $n=8$, $p=4$, $d=2$, $N=3$, $m=2$  (here $p=\frac{n}{2}$); $n=10$, $p=4$, $d=2$, $N=4$, $m=2$  (here $p=\frac{n}{2}-1$); $n=11$, $p=6 [3]$, $d=3$, $N=3$, $m=2 [1]$ etc. {Cf.~\eqref{mag_n=11_p=3} for an explicit example.}

	In particular, this always works when $p=\frac{n}{2}$ with $p$ even, i.e., when $n$ is a multiple of 4 (in this case $d=2$, $N=p-1$), see example \eqref{metric_n=8_p=4_II}.

	\item $p=\frac{n}{2}-2$ with $p$ even: these solutions can be constructed as explained for the electric forms with $p=\frac{n}{2}$ in section~\ref{subsec_elec}.

\end{itemize}

\subsection{Electric fields}

\label{subsec_elec}

\begin{itemize}

	\item $p=2$: this is the standard case studied in \cite{OrtPodZof08} (eq.~\eqref{eqE} becomes an identity) containing, e.g., Schwarzschild-Tangherlini black holes with electric charge \cite{Tangherlini63}.
	
	\item $p=3$: here $\e_{i}$ is a 1-form and this case is obviously forbidden by \eqref{eqE} (indeed dual to a magnetic field with $p=n-3$, also forbidden as already discussed), i.e., $\e_{i}=0$.

	\item $p=4$: here $\e_{ij}$ is a 2-form and (as for a magnetic field with $p=2$ \cite{OrtPodZof08} --- cf. also section~\ref{subsec_magn}) $n$ must be even and $h_{ij}$ almost-K\"ahler (almost-Hermitian if $n=6$).

	\item $p=n-2$: this is dual to a magnetic 2-form so again $n$ must be even and $h_{ij}$ almost-K\"ahler (almost-Hermitian if $n=6$). Using also the previous observations one sees, e.g., that for $n=5$ only $p=2$ is permitted, for $n=6$ only $p=2,4$, for $n=7$ only $p=2$, etc. This implies, in particular, that the $\E^2r^{3-n}\ln r$ term in $2H$ (eq.~\eqref{H_log+}) can occur only for $p=(n+1)/2\ge 5$ with $n\ge 9$ ($n$ odd).

	\item $p=\frac{n}{2}+1$ ($n$ even): in this case (dual to a magnetic field with $p=\frac{n}{2}-1$) the electric field does not depend on $r$ and {if $h_{ij}$ is a direct product of two ($\frac{n}{2}-1$)-dimensional spaces then} a possible $\e_{i_1\ldots i_{p-2}}$ can be written as ($\d u\wedge\d r)\wedge$(a sum of the volume elements of the two subspaces), cf. also \cite{BarCalCha12}; for even $p$ it is $h$-self-dual. The electric field does not contribute to $T_{ij}$. See example~\eqref{el_n=8_p=5}.
	
	\item $p-2=m'd$, $n-2=Nd$ ($d\ge2$): this is dual to an already discussed magnetic field with $p=md$ (with $m=N-m'$), and one can similarly use eq.~(3.25) of \cite{BarCalCha12}. As in the magnetic case, this always works when $p=\frac{n}{2}$ with $p$ even (with $d=2$, $N=p-1$), see example \eqref{metric_n=8_p=4_II}. (For $N=2$, $m'=1$ this reduces to the case $p=\frac{n}{2}+1$ discussed above.) {By combining the constructions for the electric and magnetic cases, for $d=2$ ($m=m'+1$) one can also construct dyonic fields \cite{BarCalCha12}.}

	\item $p=\frac{n}{2}+2$ with $p$ even: these solutions can be constructed as explained for the magnetic forms with $p=\frac{n}{2}$ in section~\ref{subsec_magn}.

\end{itemize}

\subsection{Case $2p=n$ ($n$ even)}

\label{app_comm_special}

It should be recalled that in the case $p=\frac{n}{2}$, if both an electric and a magnetic field are present then they generically obey the weaker constraint \eqref{eqEB} (the corresponding ``Maxwell'' equations are also generically modified, as discussed in section~\ref{subsec_2p=n_summary}). The construction with $p=\frac{n}{2}$  (with $p$ even) mentioned above in sections~\ref{subsec_magn} and \ref{subsec_elec} still provides examples in the special case when \eqref{eqE} and \eqref{eqB} are satisfied separately (and not just \eqref{eqEB}), but more general solutions also exist, see for example \eqref{metric_n=8_p=4_II_2}. Note, however, that in the case $n=6$, $p=3$, not only \eqref{eqE} and \eqref{eqB} (as discussed in sections~\ref{subsec_magn} and \ref{subsec_elec}) but also the weaker constraint \eqref{eqEB} can be satisfied only trivially by $\E^2=0=\B^2$, as can be seen by directly substituting into \eqref{eqEB} the most general possible form of a 1-form ${\mbox{\boldmath{$e$}}}$ and a 3-form ${\mbox{\boldmath{$b$}}}$ (examples are given by \eqref{6D_null}, \eqref{6D_null_K}). Therefore, solutions $2p=n$ for which ${\mbox{\boldmath{$e$}}}$ and ${\mbox{\boldmath{$b$}}}$ are not both zero require $n\ge8$ ($p\ge4$) (see, e.g., \eqref{metric_n=8_p=4_II}, \eqref{metric_n=8_p=4_II_2}).

\section{Cases $p=1$ and $p=n-1$}

\setcounter{equation}{0}

\label{app_limiting}

Here we analyze the special ranks $p=1$ and $p=n-1$. The results of section~\ref{subsec_Max1} apply also here and need not be repeated. Just note that there is no electric term  $\e_{i_1\ldots i_{p-2}}$ for $p=1$ (while $\B^2_{ij}=\bb_i\bb_j$), and no magnetic term $\bb_{i_1\ldots i_p}$ for $p=n-1$.

As observed in section~\ref{subsec_Rur}, the first difference appears in the Einstein equation for $R_{ij}$. Instead of \eqref{constrijr} and \eqref{eqB} [\eqref{eqE}] we now have the trace-free equations
\beqn
	& & \R_{ij}-\frac{\R}{n-2}h_{ij}=\b\left(\bb_i\bb_j-\frac{\B^2}{n-2}h_{ij}\right)  \hspace{30mm} (p=1) , \label{Rij_p=1} \\
	& & \R_{ij}-\frac{\R}{n-2}h_{ij}=-(n-2)(n-3)\b\left(\E^2_{ij}-\frac{\E^2}{n-2}h_{ij}\right) \qquad (p=n-1) , \label{Rij_p=n-1}
\eeqn
while \eqref{constrijs} remains true, so that, again, $h_{ij}=h^{1/(n-2)}\,\gamma_{ij}(x)$ (with $\det\gamma_{ij}=1$).

As noticed, the results of section~\ref{subsec_Max2} are still valid also here. In particular, we have $F_{u}=0$ for $p=1$ and $F_{ui_1\ldots i_{n-2}}=0$ for $p=n-1$, and \eqref{EB_u} are still true.

As for section~\ref{subsec_Rui}, we still obtain
\be
 \mu_{,i} =0 ,
\ee
but due to coincidence of certain powers of $r$, eqs.~\eqref{constrijri}, \eqref{eqEi} and \eqref{eqBi} are replaced by
\beqn
 & & (n-4)\left[\R_{,i}-\b(\B^2)_{,i}\right]=0 \qquad\qquad (p=1) , \label{EBi_limiting_1} \\
 & & (n-4)\left[\R_{,i}-\b(n-2)(\E^2)_{,i}\right]=0 \quad (p=n-1) . \label{EBi_limiting_n-1}
\eeqn
The above two equations in fact also follow from, respectively, \eqref{Rij_p=1} and \eqref{Rij_p=n-1} once the contracted Bianchi identity and the Maxwell equations in the geometry of $h_{ij}$ are employed.

Finally, the equation discussed in section~\ref{subsubsec_Ruu_gener} contains now an additional $r^{-2}$ term that, however, vanishes identically thanks to \eqref{EBi_limiting_1}, \eqref{EBi_limiting_n-1} (except for $n=4$, see section~\ref{subsec_limiting_n=4} below). Hence one still obtains $(n-2)\mu_{,u}=-(n-1)\mu(\ln\sqrt{h})_{,u}$ (for $n\neq 4$). As in section~\ref{subsubsec_Ruu_gener}, the latter equation together with \eqref{EB_u} implies that we can choose coordinates such as
\be
 h_{ij}=h_{ij}(x) , \qquad \mu=\mbox{const} , \qquad (\B^2)_{,u}=0 \quad (p=1) , \qquad (\E^2)_{,u}=0 \quad (p=n-1)  \qquad (n\neq 4) ,
\ee
{which with \eqref{EBi_limiting_1}, \eqref{EBi_limiting_n-1} for $n>4$ give
\be
 \R=\b\B^2+\R_0 \quad (p=1) , \qquad \R=\b(n-2)\E^2+\R_0 \quad (p=n-1) , \label{R_p=1}
\ee
where $\R_0$ is a constant (corresponding to the Ricci scalar of the transverse geometry in the vacuum limit).}

\subsection{Case $n>4$}

To summarize, for $n>4$ the metric is given by \eqref{ds_generic} with
\be
 2H=\frac{{\R_0}}{(n-2)(n-3)}-\frac{2\Lambda}{(n-1)(n-2)}\,r^2-\frac{\mu}{r^{n-3}} ,
\ee
where $\R_0$, $\Lambda$ and $\mu$ are constants, and $h_{ij}$ is a Riemannian metric satisfying \eqref{Rij_p=1} or \eqref{Rij_p=n-1} (and thus having a Ricci scalar given by \eqref{R_p=1}). The Maxwell field is given by
\beqn
 & & \bF=\bb_{i}(x)\,\d x^{i} \quad (p=1) , \\
 & & \bF=\frac{1}{(n-3)!}r^{n-4}\e_{i_1\ldots i_{n-3}}(x)\,\d u\wedge\d r\wedge\d x^{i_1}\wedge\ldots\wedge\d x^{i_{n-3}} \quad (p=n-1) ,
\eeqn
where $\bb_{i}(x)$ and $\e_{i_1\ldots i_{n-3}}(x)$ are harmonic forms in the geometry of $h_{ij}$. In contrast to the case $2\le p\le n-2$, here the Maxwell field does not enter the metric function $H$, but instead ``back-reacts'' on the transverse geometry $h_{ij}$, which thus cannot be Einstein. The Weyl type is D(bd) (since $H=H(r)$), the Maxwell type is D, and \eqref{nulldirections} are doubly aligned null directions for both the Weyl and Maxwell tensors. The metric is static (at least in regions where $H>0$). We further observe that \eqref{Rij_p=1} gives $(n-2)\R_{ij}\bb^i\bb^j=\B^2[\R+\b(n-3)\B^2]$, which is certainly non-negative if $\R\ge0$. Since $b_i$ is harmonic, this implies (see \cite{Bochner46,Bochner48}; also theorem~2.9 of \cite{YanBocbook}) that $h_{ij}$ cannot describe a compact space having $\R\ge0$ everywhere (unless, trivially, $\R=0=\B^2$).

Thanks to \eqref{Rij_p=1} and \eqref{Rij_p=n-1}, it is easy to see that any $(n-2)$-dimensional solution of the Euclidean Einstein--Maxwell theory (i.e., a metric $h_{ij}$ coupled to a 1-form $\bb_{i}(x)$ or a $(n-3)$-form $\e_{i_1\ldots i_{n-3}}(x)$) can be used to generate an $n$-dimensional Robinson--Trautman spacetime coupled to a 1-form (or a $(n-1)$-form). For example, by taking as a ``seed'' the Euclidean version of the 3D charged BTZ metric (with ``$J=0$'') \cite{BTZ} we obtain the following 5D example.

\paragraph{Example  ($n=5$, $p=1$)}

\beqn
 & & h_{ij}\d x^i\d x^j=-(\lambda\rho^2+m+\b b_x^2\ln\rho)\d\tau^2-\frac{\d\rho^2}{\lambda\rho^2+m+\b b_x^2\ln\rho}+\rho^2\d x^2 , \nonumber \\
 & & 2H=\lambda-\frac{\Lambda}{6}\,r^2-\frac{\mu}{r^2} , \qquad  \bF=\bb_x\d x ,
\eeqn
where $\lambda=\frac{{\R_0}}{6}$, $m$ and $b_x$ are constants (respectively, the cosmological constant, mass parameter and field strength of the 3D BTZ solution), and $\R=6\lambda+\b b_x^2\rho^{-2}$. Having a Lorentzian signature requires (at least for a large $\rho$) that $\lambda<0$. The dualization to $p=4$ is obvious. We observe however that the $\rho=0$ curvature singularity of the BTZ metric extends to the full 5D solution, interestingly also beyond the horizon(s). Euclideanization of this 5D solution can be used, in turn, to produce a 7D Robinson--Trautman solutions, and so on to higher odd dimensions.

\subsection{Case $n=4$}

\label{subsec_limiting_n=4}

We observe that for $n=4$ the l.h.s. of both \eqref{Rij_p=1} ($p=1$) and \eqref{Rij_p=n-1} ($p=3$) is identically zero (since $h_{ij}$ is 2-dimensional), which implies, respectively, $\bb_i=0$ and $\e_i=0$. Since we also have $F_{u}=0$ for $p=1$ and $F_{ui_1\ldots i_{n-2}}=0$ for $p=n-1$, we conclude that $\bF$ vanishes identically and one is left only with vacuum Robinson--Trautman spacetimes, obeying the standard equation ${-}\triangle{\cal R}+4\mu_{,u}+6\mu(\ln\sqrt{h})_{,u}=0$ (this contains, in particular, the Schwarzschild metric, which has $\bF=0$ and yet can describe a black hole with non-zero axionic charge in the presence of a non-zero Kalb--Ramond field --- in fact it is the only such static and asymptotically flat solution \cite{Bowicketal88}). Therefore, in 4D the only electrovac spacetimes with $\bF\neq 0$ are obtained for $p=2$, which is the well-known Einstein--Maxwell case \cite{RobTra62,Stephanibook,GriPodbook}. We observe that certain Robinson--Trautman 4D solutions with aligned $p=3$ forms (axionic black holes) have been discussed in \cite{BarCalCha12}, but these involve multiple $p$-form fields and thus do not contradict our result.

%\bibliographystyle{JHEP}
%
%%%\bibliographystyle{my_cqg}
%%
%%%\bibliographystyle{elsart-num_mio}
%%\bibliography{bibl}
%
%%\bibliographystyle{unsrt}

\providecommand{\href}[2]{#2}\begingroup\raggedright\endgroup

\bibliography{bibl}

\end{document}